\documentclass[a4paper,12pt]{article}
\usepackage[utf8]{inputenc}
\usepackage[english]{babel}
\usepackage{graphicx}
\usepackage{mathtext}
\usepackage{braket}
\usepackage{feynmp-auto}
\usepackage{indentfirst}
\usepackage{epsfig,amsmath,amsfonts,braket,fp,bbm,tikz-cd,authblk,hyperref}
\usepackage{amssymb}
\usepackage{amsmath}
\usepackage{slashed}
\usepackage{euscript}
\usepackage{bbold}
\usepackage{bm,paralist,xspace,url,relsize}
\usepackage{color,xcolor}
\usepackage{hyperref}
\usepackage{textcomp}
\usepackage{appendix}
\usepackage{titlesec}
\usepackage{blindtext}
\usepackage{cmap} 
\usepackage[left=1.5cm,right=1.5cm,top=2cm,bottom=2cm,bindingoffset=0cm]{geometry}
\PassOptionsToPackage{unicode}{hyperref}
\PassOptionsToPackage{naturalnames}{hyperref}
\usepackage{titlesec}
\titleformat{\section}{\normalfont\large\bfseries}{\thesection}{1em}{}
\titleformat{\subsection}{\normalfont\normalsize\bfseries}{\thesubsection}{1em}{}

\usepackage[active]{srcltx}

\usepackage{tikz}

\begin{document}
 \setcounter{section}{0}

\renewcommand{\contentsname}{Contents}




\author[1,2]{E.T.Akhmedov}
\author[3]{O.Diatlyk}
\affil[1]{Moscow Institute of Physics and Technology, 141700, Dolgoprudny, Russia}
\affil[2]{Institute for Theoretical and Experimental Physics, 117218, Moscow, Russia}
\affil[3]{National Research University Higher School of Economics, 101000 Moscow, Russia}

\title{\textcolor{black}{Secularly growing loop corrections in scalar wave background}}


\numberwithin{equation}{section}

\maketitle

\begin{abstract}

We consider two--dimensional Yukawa theory in the scalar wave background $\phi(t-x)$. If one takes as initial state in such a background the scalar vacuum corresponding to $\phi=0$, then loop corrections to a certain part of the Keldysh propagator, corresponding to the anomalous expectation value, grow with time. That is a signal to the fact that under the kick of the $\phi(t-x)$ wave the scalar field rolls down the effective potential from the $\phi=0$ position to the proper ground state. We show the evidence supporting these observations.

\end{abstract}

\newpage

\tableofcontents

\newpage
\section{Introduction}
The aim of quantum field theory is to find the response of a system to external perturbations: in particular, to find correlation functions or, more generically, correlations between an external influence on the
system and its backreaction on it. In classical field theory correlation functions are solutions of equations
of motion. In quantum field theory one also should take into account quantum fluctuations, i.e. calculate
loop corrections to the tree-level correlation functions.

Usually one treats quantum fluctuations using
Feynman diagrammatic technique. It implicitly assumes that external perturbations do not change the
initial state of the theory, i.e. the system remains stationary \cite{20}, \cite{16}, \cite{2}.
However, strong background fields usually take the state of the quantum field theory out of equilibrium;
in such a situation standard (stationary or Feynman) technique incorrectly describes the dynamics of the
fields. For instance, stationary approximation is violated in an expanding universe (see, e.g., \cite{2}--\cite{5}),
in strong electric fields \cite{6},\cite{7}, during the gravitational collapse \cite{8} and in a number of other situations \cite{9}--\cite{new7}. In such situations loop corrections to the tree-level correlation functions grow
with time. This indicates the breakdown of perturbation theory. Namely, every power of the small
coupling constant is accompanied by a large (growing with evolution time) factor.

These observations raise the problem
of the resummation of at least of the leading loop corrections from all loops.
Such a resummation was performed only in a limited number of cases \cite{2}--\cite{7} \textcolor{black}{and \cite{new1}--\cite{new7}}. Moreover, even in these
cases one can catch only the leading qualitative effects in the limit of long evolution period and small
coupling constant. In this respect it would be nice to find a simple but nontrivial example of a nonequilibrium field theory, in which calculations and dynamics itself are more transparent than in complex
gravitational and electromagnetic fields. In the papers \cite{12}, \cite{46}  such an example was proposed.
There the Yukawa theory in (1+1)-dimensional Minkowski spacetime,
\begin{equation}
    \label{eq:i1}
    S=\int d^{2}x\bigg[\dfrac{1}{2}\big(\partial_{\mu}\phi\big)^{2}+i\bar{\psi} \slashed{\partial}\psi-\lambda \phi\bar{\psi}\psi\bigg],
\end{equation}
was considered. Usually one quantizes this theory on the trivial background $\phi_{cl} = 0$, $\psi_{cl} = 0$
and uses the standard equilibrium approach to find scattering amplitudes. This approach is not
applicable in the presence of a strong background scalar field $\phi_{cl} \neq 0$, at least if there is a pumping of
energy into the system, which may generate an increase of the higher level populations and anomalous
quantum averages. To study such an out of equilibrium situation, we calculate correlation functions using non-equilibrium Schwinger–Keldysh
diagrammatic technique \cite{20}--\cite{16}, \cite{14}--\cite{19}.

Let us mention the main results of the paper \cite{12}. First, we assumed that there is a  scalar
field, i.e. a classical solution\footnote{Whereas the separate
paper \cite{46} considers the case of the strong scalar field backgrounds of the forms $\phi_{cl}=\dfrac{m}{\lambda}+Et$ and $\phi_{cl}=\dfrac{m}{\lambda}+Ex$.} $\phi_{cl}=\dfrac{1}{\lambda}\Phi\bigg(\dfrac{t-x}{\sqrt{2}}\bigg)$  and $\psi_{cl} = 0$.   Second,
we split each field into the sum of the “classical background” and “quantum fluctuations”: $\phi=\phi_{cl}+\phi_{q}$, $\psi=\psi_{q}$, quantized the “quantum” part and found tree-level propagators when $\phi_{q}=0$, i.e. when the scalar field was considered as non-dynamical classical background. In \cite{12} we have used two approaches -- the functional and the operator one. In the operator formalism we used the exact fermion
modes:
\begin{equation}
    \label{eq:i2}
    \hat{\psi}(u,v)=\int_{0}^{+\infty}\dfrac{dq}{2\pi}\dfrac{1}{\sqrt[4]{2}}\bigg[\hat{a}_{q}\begin{bmatrix}1 \\ \dfrac{\Phi(v)}{\sqrt{2}q} \end{bmatrix}e^{-iqu-ia(v,0)/q}+\hat{b}_{p}^{\dagger}\begin{bmatrix}1 \\ -\dfrac{\Phi(v)}{\sqrt{2}q} \end{bmatrix}e^{iqu+ia(v,0)/q}\bigg] \ ,
\end{equation}
with

$$
a(v,0) = \dfrac{1}{2}\int_{0}^{v}dy \, \Phi^{2}(y), \quad v = \frac{t-x}{\sqrt{2}}, \quad  u = \frac{t+x}{\sqrt{2}},
$$
and
\begin{equation}
    \label{eq:i3}
 \{\hat{a}_{p},\hat{a}_{q}^{\dagger}\}=2\pi \delta(p-q), \quad
\{\hat{b}_{p},\hat{b}_{q}^{\dagger}\}=2\pi \delta(p-q),
\end{equation}
and canonical commutation relations for the field operators.

Third, we explicitly found the response of the fermion field on the scalar background,
\begin{eqnarray}
    \label{eq:i4}
    \langle  \bar{\psi}\psi\rangle (t,x) \approx \dfrac{\Phi(v)}{\pi} \,\ln\bigg[\frac{\Phi(v)}{\Lambda} \bigg],
\end{eqnarray}
at the tree–level, where $\Lambda$ is the UV cutoff. And finally, we calculated the expectation value of the fermion flux:
\begin{equation}
    \label{e:i5}
   \langle T^{01} \rangle_{reg} \approx -\dfrac{1}{48\pi}\{\bar{a}(v),v\} \ , \qquad \text{where} \qquad \bar{a}(v)=\dfrac{1}{m^{2}}\int_{}^{v}dy \, \Phi^{2}(y),
\end{equation}
and $\{f(z),z\}$ is the Schwarzian derivative.

In the present paper, we make the scalar field dynamical $\phi_{q}\neq 0$ and calculate loop corrections to the correlation functions using non-equilibrium Schwinger–Keldysh diagrammatic technique. In particular, we are mostly interested in the loop corrections to the Keldysh
propagators for the scalar and fermion fields, because these propagators reflect the change of the state of the
theory (see e.g. \cite{16}, \cite{2}). Also, we calculate one loop corrections to the vertexes and 3-point scalar correlation functions. At the loop level some contributions to these quantities may receive growing with time, so called secular, loop corrections. The usual equilibrium Feynman technique is not applicable in such calculations. For instance, this is the case of strong electric \cite{6},\cite{7} and gravitational \cite{2},\cite{8} fields, where
loop corrections to the Keldysh propagator grow with time.
Feynman technique takes into account only contributions of the zero point fluctuations into correlation
functions. To take into account the change of the initial state of the theory (change in the anomalous
averages) and of the excitation of higher levels (for the exact modes in background fields)
one has to apply the Schwinger–Keldysh technique.

The goal of this paper is to find out whether a secular growth of loop corrections to different propagators and correlation functions for bosons or fermions does appear or does not. If there is such an effect, then to understand the consequences one has to resum the leading contributions from all loops with the help of Schwinger-Dyson equations.

In section \ref{Section3} we analyze one-loop corrections to the 1-,2- and 3-point correlation functions for scalars. We find that only the anomalous quantum averages for scalar field with positive momenta possess secular growth, while the level population receives subleading loop corrections. Also, we show that 3-point scalar correlation functions as well as $n$--point ones receive secularly growing corrections, which are suppressed, however, by higher powers of $\lambda$.

In section \ref{Section4} we find that loop corrections to the fermion propagators as well as to the vertexes do not receive any secularly growing loop corrections. This observation in particular means that the tree level flux \eqref{e:i5} and the scalar current \eqref{eq:i4} do not receive any large modifications in the loops.
In section \ref{Section5} we derive the Dyson-Schwinger equation for the anomalous average which sums up the leading corrections from all loops. Then, we find a stationary solution of this equation and explain its physical meaning.

We discuss the results and conclude in section \ref{Section6}. To make the paper self contained and to simplify the presentation in its main body we put some calculations into the Appendix.

\section{Action, modes and tree-level Green functions}

We consider Yukawa theory in (1+1)-dimensional Minkowski spacetime with the action:
\begin{equation}
    \label{eq:1}
   S[\psi, \bar{\psi},\phi]=  \int_{\mathcal{C}} dt\int dx \bigg( \dfrac{1}{2}\partial_{\mu}\phi \partial^{\mu}\phi+ \bar{\psi}i \slashed{\partial}\psi-\lambda \phi\bar{\psi}\psi \bigg).
\end{equation}
The signature of the metric is (1,-1) and $\mathcal{C}$ is the Keldysh closed time contour. We use the Weyl representation for the Clifford algebra:

\begin{equation}
    \label{eq:2}
    \gamma^{0}=\left[ \begin{array}{cc}
   0 & 1 \\
    1 & 0 \\
  \end{array} \right],  \qquad  \gamma^{1}=\left[ \begin{array}{cc}
   0 & 1 \\
    -1 & 0 \\
  \end{array} \right].
\end{equation}
In the presence of classical background fields we split $\psi=\psi_{cl}+\psi_{q}$ and $\phi=\phi_{cl}+\phi_{q}$, where $\phi_{cl}$, $\psi_{cl}$ are solutions of classical equations of motion:

\begin{equation}
    \label{eq:3}
    \begin{cases}
    \partial^{2}\phi_{cl} + \lambda \, \bar{\psi}_{cl} \, \psi_{cl} = 0 \\
    \big[ i \gamma^{\mu}\partial_{\mu} - \lambda \, \phi_{cl}\big]\, \psi_{cl} = 0 \ ,
    \end{cases}
\end{equation}
and $\phi_q$, $\psi_q$ are quantum fluctuations.
In particular we consider the following scalar wave solution:
\begin{equation}
    \label{eq:4}
    \lambda \, \phi_{cl}(t,x) = \Phi\bigg(\dfrac{t-x}{\sqrt{2}}\bigg) \quad {\rm and} \quad
    \psi_{cl}=0.
\end{equation}
We introduce the new pair of fields:
\begin{equation}
    \label{eq:5}
\begin{cases}
\Psi_{1}=\dfrac{1}{\sqrt{2}}\bigg(\psi_{+}+\psi_{-}\bigg) \\ \Psi_{2}=\dfrac{1}{\sqrt{2}}\bigg(\psi_{+}-\psi_{-}\bigg)
\end{cases},
\begin{cases}
\Bar{\Psi}_{1}=\dfrac{1}{\sqrt{2}}\bigg(\bar{\psi}_{+}-\bar{\psi}_{-}\bigg) \\ \Bar{\Psi}_{2}=\dfrac{1}{\sqrt{2}}\bigg(\bar{\psi}_{+}+\bar{\psi}_{-}\bigg)
\end{cases} \quad  \text{and} \qquad \begin{cases}
\phi_{1}=\dfrac{1}{\sqrt{2}}\bigg(\phi_{+}+\phi_{-}\bigg) \\ \phi_{2}=\dfrac{1}{\sqrt{2}}\bigg(\phi_{+}-\phi_{-}\bigg)
\end{cases} \ ,
    \end{equation}
where $\phi_{+}$ and $\psi_{+}$ are $\phi$ and $\psi$ on the upper part of the contour $\mathcal{C}$, while $\phi_{-}$ and $\psi_{-}$ are $\phi$ and $\psi$ on the lower one. Then the action \eqref{eq:1} acquires the following form:
\begin{equation}
    \label{eq:6}
        S[\psi, \bar{\psi},\phi]=  \int dt\int dx \bigg( \dfrac{1}{2}\partial_{\mu}\phi_{1} \partial^{\mu}\phi_{1}+\dfrac{1}{2}\partial_{\mu}\phi_{2} \partial^{\mu}\phi_{2} +\bar{\Psi}(x,t)i \slashed{\partial}\Psi(x,t)-\bar{\Psi}(x,t)\hat{\Phi}(t-x)\Psi(x,t) - V_{int}\bigg).
\end{equation}
with
\begin{equation}
    \label{eq:7}
    V_{int}=\dfrac{\lambda}{\sqrt{2}}\bigg(\phi_{1}\bar{\Psi}_{1}\Psi_{1}+\phi_{1}\bar{\Psi}_{2}\Psi_{2}+\phi_{2}\bar{\Psi}_{1}\Psi_{2}+\phi_{2}\bar{\Psi}_{2}\Psi_{1}\bigg),
\end{equation}
where in the action \eqref{eq:6} we use the following notation
\begin{equation}
    \label{eq:8}
    \Psi=\begin{bmatrix}\Psi_{1} \\ \Psi_{2} \end{bmatrix},
\quad \quad \Bar{\Psi}=\begin{bmatrix}\Bar{\Psi}_{1} \\ \Bar{\Psi}_{2} \end{bmatrix}
\quad \text{and} \quad  \hat{\Phi}(t,x)=\left[ {\begin{array}{cc}
   \Phi(t-x)& 0 \\
    0 & \Phi(t-x) \\
  \end{array} } \right].
\end{equation}
After such a rotation as \eqref{eq:5} the propagator matrix transforms into the triangular form:
\begin{equation}
    \label{eq:9}
   G(t,x;t^{\prime},x^{\prime})=
  \left[ {\begin{array}{cc}
   -i\braket{\Psi_{1}(t,x)\bar{\Psi}_{1}(t^{\prime},x^{\prime})} & -i\braket{\Psi_{1}(t,x)\bar{\Psi}_{2}(t^{\prime},x^{\prime})} \\
   0 & -i\braket{\Psi_{2}(t,x)\bar{\Psi}_{2
  }(t^{\prime},x^{\prime})} \\
  \end{array} } \right]=\left[ {\begin{array}{cc}
   G^{R} & G^{K} \\
    0 & G^{A} \\
  \end{array} } \right],
\end{equation}
and the scalar propagator matrix becomes:
\begin{equation}
    \label{eq:10}
   D(t,x;t^{\prime},x^{\prime})=
  \left[ {\begin{array}{cc}
   -i\braket{\phi_{1}(t,x)\phi_{1}(t^{\prime},x^{\prime})} & -i\braket{\phi_{1}(t,x)\phi_{2}(t^{\prime},x^{\prime})} \\
   -i\braket{\phi_{2}(t,x)\phi_{1
  }(t^{\prime},x^{\prime})} & 0 \\
  \end{array} } \right]=\left[ {\begin{array}{cc}
   D^{K} & D^{R} \\
    D^{A} & 0 \\
  \end{array} } \right].
\end{equation}
Furthermore, the field operator for the free massless scalar field has the form:
\begin{equation}
    \label{eq:11}
    \hat{\phi}(t,x)=\int_{-\infty}^{+\infty}\dfrac{dp}{2\pi}\bigg(\hat{\alpha}_{p}f_{p}(t,x)+\hat{\alpha}_{p}^{\dagger}f^{*}_{p}(t,x)\bigg),
\end{equation}
where $f_{p}(t,x)=\dfrac{1}{\sqrt{2|p|}}e^{-i|p|t+ipx}$ and operators $\hat{\alpha}_{p}$, $\hat{\alpha}_{p}^{\dagger}$  satisfy the standard commutation relations: $[\hat{\alpha}_{p},\hat{\alpha}_{q}^{\dagger}]=2\pi\delta(p-q)$.
The tree-level Keldysh propagator for the scalar field is
\begin{equation}
    \label{eq:12}
    D^{K}_{0}(t,x;t^{\prime},x^{\prime})=-i\langle0| \{\hat{\phi}(t,x),\hat{\phi}(t^{\prime},x^{\prime})\}|0\rangle=-i\int_{-\infty}^{+\infty} \dfrac{dp}{2\pi}\bigg(f_{p}(t,x)f^{*}_{p}(t^{\prime},x^{\prime})+f_{p}^{*}(t,x)f_{p}(t^{\prime},x^{\prime})\bigg),
\end{equation}
where $\hat{\alpha}_{p}|0\rangle=0$. At the same time the retarded propagator for the scalar field is:
\begin{equation}
    \label{eq:13}
    D^{R}_{0}(t,x;t^{\prime},x^{\prime})=-i\theta(t-t^{\prime}) [\hat{\phi}(t,x),\hat{\phi}(t^{\prime},x^{\prime})]=-i\theta(t-t^{\prime})\int_{-\infty}^{+\infty} \dfrac{dp}{2\pi}\bigg(f_{p}(t,x)f^{*}_{p}(t^{\prime},x^{\prime})-f^{*}_{p}(t,x)f_{p}(t^{\prime},x^{\prime})\bigg),
\end{equation}
and is state independent.
The advanced propagator is conjugate to the retarded one:
\begin{equation}
     D^{R}_{0}(1;2)=D^{A}_{0}(2;1).
\end{equation}
The field operator for fermions\footnote{The form of the field operator under consideration is formal, because of the peculiarity of the modes at $p=0$. That is why in the calculations of observables we assume that there is an $\epsilon$ shift in the exponents: $u \rightarrow u - i \epsilon, \quad v \rightarrow v -i\epsilon.$} has the following mode expansion \cite{12}:

\begin{equation}
    \label{eq:15}
    \hat{\psi}(u,v)=\int_{0}^{+\infty}\dfrac{dq}{2\pi}\dfrac{1}{\sqrt[4]{2}}\bigg[\hat{a}_{q}u_{p}(v)e^{-iqu-ia(v,0)/q}+\hat{b}_{p}^{\dagger}v_{p}(v)e^{iqu+ia(v,0)/q}\bigg] \ ,
\end{equation}
with
\begin{equation}
    \label{eq:16}
 \{\hat{a}_{p},\hat{a}_{q}^{\dagger}\}=2\pi \delta(p-q), \quad
\{\hat{b}_{p},\hat{b}_{q}^{\dagger}\}=2\pi \delta(p-q),
\end{equation}
and
\begin{equation}
    \label{eq:17}
    u_{p}(v)=\begin{bmatrix}1 \\ \dfrac{\Phi(v)}{\sqrt{2}q} \end{bmatrix}, \qquad v_{p}(v)=\begin{bmatrix}1 \\ -\dfrac{\Phi(v)}{\sqrt{2}q} \end{bmatrix}, \qquad \text{where} \qquad a(v,0)=\int_{0}^{v}\Phi^{2}(y)dy,
\end{equation}
where we have introduced the light-cone coordinates
\begin{equation}
    u=\dfrac{t+x}{\sqrt{2}} \qquad \text{and} \qquad v=\dfrac{t-x}{\sqrt{2}}.
\end{equation}
Also we use the following notation for the modes:
\begin{equation}
    \label{equation16}
    \chi_{p}=\dfrac{1}{\sqrt[4]{2}}u_{p}(v)e^{-iqu-ia(v,0)/q}  \qquad \text{and}\qquad \zeta_{p}=\dfrac{1}{\sqrt[4]{2}}v_{p}(v)e^{iqu+ia(v,0)/q}.
\end{equation}
Then the tree-level Keldysh propagator for the fermionic field is:
\begin{equation}
    \label{eq:18}
    G_{0}^{K}(t,x;t^{\prime},x^{\prime})=-\dfrac{i}{\sqrt{2}}\int_{0}^{+\infty}\dfrac{dp}{2\pi}\bigg(u_{p}(v)\bar{u}_{p}(v^{\prime})e^{-ip(u-u^{\prime})-ia(v,v^{\prime})/p}-v_{p}(v)\bar{v}_{p}(v^{\prime})e^{ip(u-u^{\prime})+ia(v,v^{\prime})/p}\bigg),
\end{equation}
when calculated for the state $\hat{a}_{p}|0\rangle=\hat{b}_{p}|0\rangle=0$. The retarded propagator is
\begin{equation}
    \label{eq:19}
    G_{0}^{R}(t,x;t^{\prime},x^{\prime})=-\dfrac{i}{\sqrt{2}}\theta(t-t^{\prime})\int_{0}^{+\infty}\dfrac{dp}{2\pi}\bigg(u_{p}(v)\bar{u}_{p}(v^{\prime})e^{-ip(u-u^{\prime})-ia(v,v^{\prime})/p}+v_{p}(v)\bar{v}_{p}(v^{\prime})e^{ip(u-u^{\prime})+ia(v,v^{\prime})/p}\bigg),
\end{equation}
and is state independent.
The advanced propagator is
\begin{equation}
    \label{eq:20}
    G_{0}^{A}(t,x;t^{\prime},x^{\prime})=\dfrac{i}{\sqrt{2}}\theta(t^{\prime}-t)\int_{0}^{+\infty}\dfrac{dp}{2\pi}\bigg(u_{p}(v)\bar{u}_{p}(v^{\prime})e^{-ip(u-u^{\prime})-ia(v,v^{\prime})/p}+v_{p}(v)\bar{v}_{p}(v^{\prime})e^{ip(u-u^{\prime})+ia(v,v^{\prime})/p}\bigg).
\end{equation}
where we denoted
\begin{equation}
    \label{eq:21}
    u_{p}(v)\bar{u}_{p}(v^{\prime})=\left[ {\begin{array}{cc}
   \dfrac{\Phi(v^{\prime})}{\sqrt{2}p}& 1 \\
    \dfrac{\Phi(v)\Phi(v^{\prime})}{2p^{2}} & \dfrac{\Phi(v)}{\sqrt{2}p} \\
  \end{array} } \right] \qquad \text{and} \qquad  v_{p}(v)\bar{v}_{p}(v^{\prime})=\left[ {\begin{array}{cc}
   -\dfrac{\Phi(v^{\prime})}{\sqrt{2}p}& 1 \\
    \dfrac{\Phi(v)\Phi(v^{\prime})}{2p^{2}} & -\dfrac{\Phi(v)}{\sqrt{2}p} \\
  \end{array} } \right].
\end{equation}

\section{ Loop corrections to the boson correlation functions}\label{Section3}

In this section we calculate one loop corrections to the following correlation functions\footnote{Due to their causal structure the retarded and advanced propagators do not receive growing
with time quantum loop correction  in the limit $\dfrac{t_{1}+t_{2}}{2} \gg |t_{1}-t_{2}|$ (see e.g. \cite{16}, \cite{46}).  }:
\begin{equation}
    \label{i7}
    \begin{cases}
    1. \ \langle \phi_{i}(t,x) \rangle \qquad \text{for} \qquad i=1,2 \\
    2. \ D^{K}(t_{1},x_{1};t_{2},x_{2}) \\
    3. \ \langle \phi_{i}(t_{1},x_{1})\phi_{j}(t_{2},x_{2})\phi_{k} (t_{3},x_{3})\rangle \qquad \text{for} \qquad i,j,k=1,2
    \end{cases}
\end{equation}
in the limit $t_{1} \approx t_{2} \approx t_{3} \approx T \rightarrow \infty$. Also, we assume that evolution of the system starts after a moment of time $T_{0}\rightarrow -\infty$, i.e. we set up the initial state $\hat{\alpha}_{p}|0\rangle=\hat{a}_{p}|0\rangle=\hat{b}_{p}|0\rangle=0$ at this moment and adiabatically turn on interactions, $\lambda$, after it.

We show below that:

1. it is possible to interpret the one loop correction to the one point correlation function in terms of the effective action for scalars which was found in \cite{46};

2. the anomalous average $\kappa_{pq} \equiv \langle\hat{\alpha}_{p}\hat{\alpha}_{q} \rangle $, which is an element of the boson's Keldysh propagator as we show below, receives a secularly growing loop correction ($\sim \lambda^{2}(T-T_{0})$) only if both $p,q$ are positive, while the expectation value of the level population $n_{pq}\equiv \langle\hat{\alpha}^{\dagger}_{p}\hat{\alpha}_{q} \rangle $ does not receive growing with time corrections at all;

3. one loop corrections to the three-point correlation functions also receives growing with time loop corrections, but they are suppressed by higher power of  $\lambda$  ($\sim \lambda^{3}(T-T_{0})$);

\subsection{One loop corrections to the one point scalar correlation functions}

Using $V_{int}$ from \eqref{eq:7} and the tree-level propagators defined in previous section, we get that the loop corrections to the one point correlation function has the following form:
\begin{eqnarray}
    \label{eq:102}
    \langle \phi_{1}(t,x) \rangle= \nonumber \\
    =\dfrac{-i\lambda}{\sqrt{2}}(i)^{2}\int d^{2}x_{1}\bigg[D_{0}^{K}(t,x;t_{1},x_{1})tr\big(G_{0}^{R}(1;1)+G_{0}^{A}(1;1)\big)+D_{0}^{R}(t,x;t_{1},x_{1})trG_{0}^{K}(1;1)\bigg]+o(\lambda^{3}). \ \
\end{eqnarray}
Using the results of \cite{12}
$$trG_{0}^{K}(1;1)=2i\dfrac{\Phi(v_{1})}{\pi}\ln{\bigg(\dfrac{\Phi(v_{1})}{\Lambda}\bigg)}, $$ $$tr\bigg[G_{0}^{R}(1;1)+G_{0}^{A}(1;1)\bigg]=0,$$
where the trace is taken over the spinor indexes and the fact that
$$D_{0}^{R}(t,x;t_{1},x_{1})=\dfrac{1}{2}\theta(u-u_{1})\theta(v-v_{1}),$$
we get that
\begin{eqnarray}
    \label{eq:103}
    \langle \phi_{1}(t,x) \rangle
    =-\dfrac{\lambda}{\sqrt{2}}\int_{}^{u} du_{1}\int_{}^{v} dv_{1}\dfrac{\Phi(v_{1})}{\pi}\ln{\bigg(\dfrac{\Phi(v_{1})}{\Lambda}\bigg)}+o(\lambda^{3}).
\end{eqnarray}
Similarly\footnote{It is, probably, worth mentioning here that the difference between $\langle \phi_{1} \rangle$ and $\langle \phi_{2} \rangle$ should be present because $\langle\phi_{2}\phi_{2} \rangle=0$.},
\begin{eqnarray}
    \label{eq:104}
    \langle \phi_{2}(t,x) \rangle= \nonumber \\
    =\dfrac{-i\lambda}{\sqrt{2}}(i)^{2}\int d^{2}x_{1}\bigg[D_{0}^{A}(t,x;t_{1},x_{1})tr\big(G_{0}^{R}(1;1)+G_{0}^{A}(1;1)\big)\bigg]+o(\lambda^{3})=o(\lambda^{3}),
\end{eqnarray}
i.e. $\langle \phi_{2}(t,x) \rangle \ll \langle \phi_{1}(t,x) \rangle$. To have a better understanding of equation \eqref{eq:103} we can consider field $\phi$ which is located on the upper branch of the Keldysh contour \eqref{eq:5}:
\begin{equation}
    \label{opp1}
    \phi_{+}=\dfrac{\phi_{1}+\phi_{2}}{\sqrt{2}}.
\end{equation}
Then, from \eqref{eq:103} and \eqref{eq:104} it follows that
\begin{equation}
    \label{eq:opp2}
    \langle \phi_{+} \rangle \approx -\dfrac{\lambda}{2}\int_{}^{u} du_{1}\int_{}^{v} dv_{1}\dfrac{\Phi(v_{1})}{\pi}\ln{\bigg(\dfrac{\Phi(v_{1})}{\Lambda}\bigg)}
\end{equation}
or equivalently
\begin{equation}
    \label{eq:opp3}
   \partial^{2}\langle \phi_{+} \rangle+\lambda\dfrac{\Phi(v)}{\pi}\ln{\bigg(\dfrac{\Phi(v)}{\Lambda}\bigg)} \approx 0,
\end{equation}
where $\partial^{2}=2\partial_{u}\partial_{v}$. We can get this equation by iteratively solving the equation obtained in \cite{46}:
\begin{equation}
    \label{eq:opp4}
    \partial^{2}\langle \phi_{+} \rangle+\lambda\dfrac{\langle \phi_{+} \rangle}{\pi}\ln{\bigg(\dfrac{\langle \phi_{+} \rangle}{\Lambda}\bigg)}=0.
\end{equation}
Note that $\partial^{2}\Phi(v)=0$ and $\dfrac{\lambda \Phi(v)}{\pi}\ln\dfrac{\Phi(v)}{\Lambda}=\dfrac{\partial V_{eff}}{\partial \Phi}$, where $V_{eff}$ is the effective scalar potential generated by quantum fluctuations of fermions in the background field in question.

\subsection{One loop corrections to the Keldysh propagator for scalars}

\begin{figure}
    \centering
    \includegraphics[width=7cm]{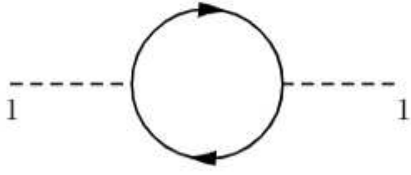}
    \caption{One loop correction to the boson Keldysh propagator. Solid lines correspond to the tree-level fermion propagators, dashed lines correspond to the tree-level boson propagators.}
    \label{fig1}
\end{figure}

The next step is to calculate
the first loop correction to the Keldysh propagator for the scalar field:
 \begin{eqnarray}
     \label{eq:22}
      \Delta_{2} D^{K}(t_{1},x_{1};t_{2},x_{2})=\dfrac{\lambda^{2}}{2}\int d^{2}x_{3}d^{2}x_{4} \times \nonumber \\
     \times \bigg[
     D_{0}^{K}(t_{1},x_{1};t_{3},x_{3})G_{0}^{K}(t_{3},x_{3};t_{4},x_{4})G_{0}^{R}(t_{4},x_{4};t_{3},x_{3})D_{0}^{A}(t_{4},x_{4};t_{2},x_{2})+ \nonumber \\
      +D_{0}^{K}(t_{1},x_{1};t_{3},x_{3})G_{0}^{A}(t_{3},x_{3};t_{4},x_{4})G_{0}^{K}(t_{4},x_{4};t_{3},x_{3})D_{0}^{A}(t_{4},x_{4};t_{2},x_{2})+ \nonumber \\
      +D_{0}^{R}(t_{1},x_{1};t_{3},x_{3})G_{0}^{R}(t_{3},x_{3};t_{4},x_{4})G_{0}^{K}(t_{4},x_{4};t_{3},x_{3})D_{0}^{K}(t_{4},x_{4};t_{2},x_{2})+ \nonumber \\
      +D_{0}^{R}(t_{1},x_{1};t_{3},x_{3})G_{0}^{K}(t_{3},x_{3};t_{4},x_{4})G_{0}^{A}(t_{4},x_{4};t_{3},x_{3})D_{0}^{K}(t_{4},x_{4};t_{2},x_{2})+ \nonumber \\
      +D_{0}^{R}(t_{1},x_{1};t_{3},x_{3})G_{0}^{R}(t_{3},x_{3};t_{4},x_{4})G_{0}^{A}(t_{4},x_{4};t_{3},x_{3})D_{0}^{A}(t_{4},x_{4};t_{2},x_{2})+ \nonumber \\
      +D_{0}^{R}(t_{1},x_{1};t_{3},x_{3})G_{0}^{K}(t_{3},x_{3};t_{4},x_{4})G_{0}^{K}(t_{4},x_{4};t_{3},x_{3})D_{0}^{A}(t_{4},x_{4};t_{2},x_{2})+ \nonumber \\
      +D_{0}^{R}(t_{1},x_{1};t_{3},x_{3})G_{0}^{A}(t_{3},x_{3};t_{4},x_{4})G_{0}^{R}(t_{4},x_{4};t_{3},x_{3})D_{0}^{A}(t_{4},x_{4};t_{2},x_{2})\bigg],
 \end{eqnarray}
which is given by a sum of diagrams of the type depicted on the fig.\ref{fig1}.

Since at any loop order the Keldysh
propagator can be represented as\footnote{Note that from now on we frequently use the notation $f_p(1) = f_p(t_1, x_1)$ and $\chi_p(1) = \chi_p(t_1,x_1)$ and etc..}:
\begin{eqnarray}
    \label{eq:23}
   D_{exact}^{K}(t_{1},x_{1};t_{2},x_{2})=-i \int \dfrac{dp}{2\pi}\int \dfrac{dq}{2\pi} \bigg[ \big(2\pi \delta (p-q)+2n_{qp}\big)f_{p}(1)f_{q}^{*}(2)+2\kappa_{pq}f_{p}(1)f_{q}(2) + h.c.\bigg], \ \
\end{eqnarray}
we can analyze the contributions to the Keldysh propagtor in terms of those to $n_{qp}$ and $\kappa_{pq}$ in the limit $\dfrac{t_{1}+t_{2}}{2} \gg t_{1}-t_{2}$:
\begin{eqnarray}
    \label{2}
    2n^{(2)}_{qp}=\dfrac{\lambda^{2}}{2}\int d^{2}x_{3}d^{2}x_{4} \times \nonumber \\
    \bigg[i\theta\big(t_{2}-t_{4}\big)f^{*}_{p}(3)f_{q}(4)G_{0}^{K}(3;4)G_{0}^{R}(4;3)+ \nonumber \\
      +i\theta\big(t_{2}-t_{4}\big)f^{*}_{p}(3)f_{q}(4)G_{0}^{A}(3;4)G_{0}^{K}(4;3)+ \nonumber \\
      -i\theta\big(t_{1}-t_{3}\big)f^{*}_{p}(3)f_{q}(4)G_{0}^{R}(3;4)G_{0}^{K}(4;3)+ \nonumber \\
      -i\theta\big(t_{1}-t_{3}\big)f^{*}_{p}(3)f_{q}(4)G_{0}^{K}(3;4)G_{0}^{A}(4;3)+ \nonumber \\
      +i\theta\big(t_{1}-t_{3}\big)\theta\big(t_{2}-t_{4}\big)f^{*}_{p}(3)f_{q}(4)G_{0}^{R}(3;4)G_{0}^{A}(4;3)+ \nonumber \\
      +i\theta\big(t_{1}-t_{3}\big)\theta\big(t_{2}-t_{4}\big)f^{*}_{p}(3)f_{q}(4)G_{0}^{K}(3;4)G_{0}^{K}(4;3)+ \nonumber \\
      +i\theta\big(t_{1}-t_{3}\big)\theta\big(t_{2}-t_{4}\big)f^{*}_{p}(3)f_{q}(4)G_{0}^{A}(3;4)G_{0}^{R}(4;3)\bigg],
\end{eqnarray}
and
\begin{eqnarray}
    \label{3}
    2\kappa^{(2)}_{pq}=\dfrac{\lambda^{2}}{2}\int d^{2}x_{3}d^{2}x_{4} \times \nonumber \\
    \bigg[(-1)i\theta\big(t_{2}-t_{4}\big)f^{*}_{p}(3)f^{*}_{q}(4)G_{0}^{K}(3;4)G_{0}^{R}(4;3)+ \nonumber \\
      +(-1)i\theta\big(t_{2}-t_{4}\big)f^{*}_{p}(3)f^{*}_{q}(4)G_{0}^{A}(3;4)G_{0}^{K}(4;3)+ \nonumber \\
      (+1)(-i)\theta\big(t_{1}-t_{3}\big)f^{*}_{p}(3)f^{*}_{q}(4)G_{0}^{R}(3;4)G_{0}^{K}(4;3)+ \nonumber \\
      (+1)(-i)\theta\big(t_{1}-t_{3}\big)f^{*}_{p}(3)f^{*}_{q}(4)G_{0}^{K}(3;4)G_{0}^{A}(4;3)+ \nonumber \\
      +(-1)i\theta\big(t_{1}-t_{3}\big)\theta\big(t_{2}-t_{4}\big)f^{*}_{p}(3)f^{*}_{q}(4)G_{0}^{R}(3;4)G_{0}^{A}(4;3)+ \nonumber \\
      +(-1)i\theta\big(t_{1}-t_{3}\big)\theta\big(t_{2}-t_{4}\big)f^{*}_{p}(3)f^{*}_{q}(4)G_{0}^{K}(3;4)G_{0}^{K}(4;3)+ \nonumber \\
      +(-1)i\theta\big(t_{1}-t_{3}\big)\theta\big(t_{2}-t_{4}\big)f^{*}_{p}(3)f^{*}_{q}(4)G_{0}^{A}(3;4)G_{0}^{R}(4;3)\bigg],
\end{eqnarray}
where the upper script $2$ means that we calculated these quantities at the $\lambda^{2}$ order.
Using \eqref{eq:18}-\eqref{eq:21} we get:
\begin{eqnarray}
    \label{eq:25}
    2n^{(2)}_{qp}=\dfrac{i}{\sqrt{2|p|2|q|}}\int_{0}^{+\infty}\dfrac{dk_{1}}{2\pi}\int_{0}^{+\infty}\dfrac{dk_{2}}{2\pi}\dfrac{1}{2}\bigg(\dfrac{k_{1}-k_{2}}{k_{1}k_{2}}\bigg)^{2}\int d^{2}x_{3}\int d^{2}x_{4}\Phi(v_{3})\Phi(v_{4}) \times\nonumber \\
 \times \bigg[ e^{-i(u_{3}-u_{4})(k_{1}+k_{1})-ia(3,4)(1/k_{1}+1/k_{2})+iu_{3}c^{-}_{p}+iv_{3}c^{+}_{p}-iu_{4}c^{-}_{q}-iv_{4}c^{+}_{q}}\bigg(-\theta_{24}\theta_{43}-\theta_{13}\theta_{34}+\theta_{13}\theta_{24}\bigg)+ \nonumber \\
    +e^{i(u_{3}-u_{4})(k_{1}+k_{1})+ia(3,4)(1/k_{1}+1/k_{2})+iu_{3}c^{-}_{p}+iv_{3}c^{+}_{p}-iu_{4}c^{-}_{q}-iv_{4}c^{+}_{q}}\bigg(\theta_{24}\theta_{43}+\theta_{13}\theta_{34}+\theta_{13}\theta_{24}\bigg)\bigg],
\end{eqnarray}
where for simplicity we denote $\theta_{ij} \equiv \theta(t_{i}-t_{j})$.
Analyzing the structure of \eqref{2} and \eqref{3} it is obvious that we can get $\kappa^{(2)}_{pq}$ if in $n^{(2)}_{qp}$ we change the overall sign and change the signs of the terms which are proportional to $\theta_{13}\theta_{34}$:
\begin{eqnarray}
    \label{10}
    2\kappa^{(2)}_{pq}=\dfrac{-i}{\sqrt{2|p|2|q|}}\int_{0}^{+\infty}\dfrac{dk_{1}}{2\pi}\int_{0}^{+\infty}\dfrac{dk_{2}}{2\pi}\dfrac{1}{2}\bigg(\dfrac{k_{1}-k_{2}}{k_{1}k_{2}}\bigg)^{2}\int d^{2}x_{3}\int d^{2}x_{4}\Phi(v_{3})\Phi(v_{4}) \times
    \nonumber \\
    \times \bigg[ e^{-i(u_{3}-u_{4})(k_{1}+k_{1})-ia(3,4)(1/k_{1}+1/k_{2})+iu_{3}c^{-}_{p}+iv_{3}c^{+}_{p}+iu_{4}c^{-}_{q}+iv_{4}c^{+}_{q}}\bigg(-\theta_{24}\theta_{43}+\theta_{13}\theta_{34}+\theta_{13}\theta_{24}\bigg)+
     \nonumber \\
    +e^{i(u_{3}-u_{4})(k_{1}+k_{1})+ia(3,4)(1/k_{1}+1/k_{2})+iu_{3}c^{-}_{p}+iv_{3}c^{+}_{p}+iu_{4}c^{-}_{q}+iv_{4}c^{+}_{q}}\bigg(\theta_{24}\theta_{43}-\theta_{13}\theta_{34}+\theta_{13}\theta_{24}\bigg)\bigg],
\end{eqnarray}
where $c^{\pm}_{p} \equiv  \dfrac{1}{\sqrt{2}}\big(|p| \pm p\big)$. In the Appendix \ref{A} we perform the calculation of the last two expressions for $n_{qp}^{(2)}$ and $\kappa^{(2)}_{pq}$. The result of the calculation in the limit $T=\dfrac{t_{1}+t_{2}}{2} \gg |t_{1}-t_{2}|$ is as follows\footnote{In the Appendix \ref{C} we make some comments of those calculations in the case $\Phi=m=const$.}:
\begin{equation}
    \label{tpp3}
    \begin{cases}
    n^{(2)}_{qp} \sim 0 \\
    \kappa^{(2)}_{pq} \approx \bigg(T-T_{0}\bigg)F(p,q)\theta(p)\theta(q)
    \end{cases},
\end{equation}
where $\theta(x)$ is the Heaviside function and
\begin{eqnarray}
    \label{eq:tpp4}
    F(p,q)\equiv-
      \dfrac{1}{\sqrt{2|p|}}\dfrac{1}{\sqrt{2|q|}}\int_{0}^{+\infty}\dfrac{dk_{1}}{2\pi}\int_{0}^{+\infty}\dfrac{dk_{2}}{2\pi}\dfrac{1}{2}\bigg(\dfrac{k_{1}-k_{2}}{k_{1}k_{2}}\bigg)^{2}\int dv_{3}\int dv_{4}\Phi(v_{3})\Phi(v_{4})\bigg[\nonumber \\
    \mathcal{P}\dfrac{1}{k_{1}+k_{2}}\bigg[e^{iv_{3}(\sqrt{2}p+(k_{1}+k_{2}))+iv_{4}(\sqrt{2}q-(k_{1}+k_{2}))+ia(v_{3},v_{4})(1/k_{1}+1/k_{2})}+ \nonumber \\
    +e^{iv_{3}(\sqrt{2}p-(k_{1}+k_{2}))+iv_{4}(\sqrt{2}q+(k_{1}+k_{2}))-ia(v_{3},v_{4})(1/k_{1}+1/k_{2})}\bigg].
\end{eqnarray}
The growth of the anomalous average $\langle\hat{\alpha}_{p}\hat{\alpha}_{q}\rangle=\kappa_{pq}$ is a sign that the initial Fock space ground state is not a proper vacuum state in the theory: field $\phi$ should roll down to the proper vacuum of the effective potential \eqref{eq:opp4}.

Another interesting observation is that anomalous average is growing only for positive momenta, i.e. for the modes of the form
\begin{equation}
    \label{eq:tpp6}
    f_{p}(t,x)=\dfrac{1}{\sqrt{2p}}e^{-i\sqrt{2}pv}.
\end{equation}
These are the modes, which are propagating in the same direction as the external field $\Phi(v)$. In concluding section we propose an explanation for these observations.

\subsection{One loop corrections to the
three-point correlation functions}\label{3.3}

\begin{figure}
    \centering
    \includegraphics[width=7.5cm]{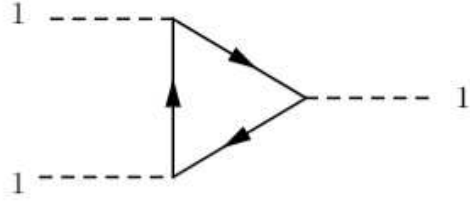}
    \caption{One loop correction to the three-point correlation function.}
    \label{fig2}
\end{figure}

In this subsection we analyze the structure of one loop corrections to the three point function. For concreteness let us consider the correlation function of the form $\langle \phi_{1}(t_{1},x_{1})\phi_{1}(t_{2},x_{2})\phi_{1}(t_{3},x_{3}) \rangle$ (fig. \ref{fig2}).
In analogy with \eqref{eq:23} one can represent the exact form of this correlation function as follows:
\begin{eqnarray}
    \label{TP:2}
  \langle\phi_{1}(t_{1},x_{1})\phi_{1}(t_{2},x_{2})\phi_{1}(t_{3},x_{3}) \rangle=\int_{-\infty}^{+\infty}\dfrac{dp}{2\pi}\dfrac{dq}{2\pi}\dfrac{dr}{2\pi}\bigg[n^{(1)}_{pqr}f_{p}(1)f_{q}(2)f_{r}(3)+n^{(2)}_{pqr}f^{*}_{p}(1)f_{q}(2)f_{r}(3)+\nonumber \\
    +n^{(3)}_{pqr}f_{p}(1)f^{*}_{q}(2)f_{r}(3)+n^{(4)}_{pqr}f_{p}(1)f_{q}(2)f^{*}_{r}(3)+h.c.\bigg].
\end{eqnarray}
Here the upper indexes of $n$'s have nothing to do with the power of $\lambda$. And again we assume that $t_{1}\approx t_{2} \approx t_{3}=T \gg |t_{i}-t_{j}|$, $i,j,k=1,2,3$.



It is interesting to point out the following observation. In Appendix \ref{A} we find that (see the previous subsection)
\begin{equation}
    \label{tpp5}
    \begin{cases}
    n^{(2)}_{qp}\propto \lambda^{2}\delta(c^{-}_{p}-c^{-}_{q})\delta \big(c^{-}_{p}+k_{1}+k_{2}\big) \\
    \kappa^{(2)}_{pq}\propto \lambda^{2} \delta(c^{-}_{p}+c^{-}_{q}) \mathcal{P}\dfrac{1}{c^{-}_{p} \pm(k_{1}+k_{2})}
    \end{cases} , \qquad \text{as} \ \ T\rightarrow +\infty \ \  \text{and} \ \  T_{0}\rightarrow -\infty,
\end{equation}
where $c^{\pm}_{p}=\dfrac{1}{\sqrt{2}}(|p|\pm p)$ and these expressions stand under the integrals over fermionic momenta $k_{1}$ and $k_{2}$ on the interval $(0,+\infty)$. Moreover, exactly these delta functions forbid the secular growth of $n_{qp}$, while allowing it for $\kappa_{pq}$: the terms multiplying these delta functions are independent of $T$ in the limit under consideration. This is because when $p,q$ are positive $c^{-}_{p} \propto p-p=0$ the expression $2\pi\delta(0)=T-T_{0}$ brings exactly the secularly growing factor into $\kappa_{pq}$. Then, even without any calculations of the scalar three point function we can make a guess that one loop corrections behave as:

$$n^{(i)}_{pqr} \propto \lambda^{3}\delta(\pm c^{-}_{p}\pm c^{-}_{q}\pm c^{-}_{r}),$$
where the sign of $c^{-}_{p,q,r}$ depends on whether the momentum $p,q$ or $r$ comes from the mode $f$ or from its complex conjugates $f^{*}$. In its turn it follows that, potentially, all of the aforementioned expectation values will contain linearly growing corrections for positive momenta $p,q,r$.

Let us calculate now the function $n^{(1)}_{pqr}$ in the approximation under consideration:
\begin{eqnarray}
    \label{TP:3}
    n^{(1)}_{pqr} \propto \lambda^{3} \int d^{2}x_{4}d^{2}x_{5}d^{2}x_{6}\bigg(f^{*}_{p}(4)f^{*}_{q}(5)f^{*}_{r}(6)+f^{*}_{p}(5)f^{*}_{q}(6)f^{*}_{r}(4)+f^{*}_{p}(6)f^{*}_{q}(4)f^{*}_{r}(5)\bigg) \times \nonumber \\
    \times \bigg[\theta(T-t_{4})tr\big\{G_{0}^{R}(4,5)G_{0}^{K}(5,6)G_{0}^{A}(6,4)\big\}+\theta(T-t_{6})tr\big\{G_{0}^{R}(4,5)G_{0}^{K}(5,6)G_{0}^{R}(6,4)\big\}+ \nonumber \\
    +\theta(T-t_{5})tr\big\{G_{0}^{A}(4,5)G_{0}^{K}(5,6)G_{0}^{A}(6,4)\big\}+\theta(T-t_{6})\theta(T-t_{4})tr\big\{G_{0}^{K}(4,5)G_{0}^{A}(5,6)G_{0}^{K}(6,4)\big\}+\nonumber \\
    +\theta(T-t_{5})\theta(T-t_{6})\bigg(tr\big\{G_{0}^{R}(4,5)G_{0}^{A}(5,6)G_{0}^{R}(6,4)\big\}+tr\big\{G_{0}^{A}(4,5)G_{0}^{R}(5,6)G_{0}^{A}(6,4)\big\}\bigg)+\nonumber \\
    +\theta(T-t_{4})\theta(T-t_{5})tr\big\{G_{0}^{K}(4,5)G_{0}^{R}(5,6)G_{0}^{K}(6,4)\big\}+\theta(T-t_{4})\theta(T-t_{5})\theta(T-t_{6})\times \nonumber \\
    \times\bigg(tr\big\{G_{0}^{A}(4,5)G_{0}^{K}(5,6)G_{0}^{R}(6,4)\big\}+\dfrac{1}{3}tr\big\{G_{0}^{K}(4,5)G_{0}^{K}(5,6)G_{0}^{K}(6,4)\big\}\bigg)\bigg]. \ \
\end{eqnarray}
First, using \eqref{eq:18}, \eqref{eq:19} and \eqref{eq:20} one can straightforwardly show that $n^{(1)}_{pqr}$ contains terms which are proportional to

$$e^{ic^{-}_{p}u_{4}+ic^{-}_{q}u_{5}+ic^{-}_{r}u_{6}+iak_{1}(u_{4}-u_{5})+ibk_{2}(u_{5}-u_{6})+ick_{3}(u_{6}-u_{4})}, \qquad \text{where} \qquad a,b,c=\pm1.$$ Also, these exponents are multiplied by Heaviside $\theta$ functions. To understand the structure of such terms let us calculate one of them: the one in \eqref{TP:3} which is proportional to the $\theta(T-t_{4})$ leads to:
\begin{eqnarray}
    \label{TP:4}
    \int du_{4}du_{5}du_{6}\theta(T-t_{4})\theta(t_{4}-t_{5})\theta(t_{4}-t_{6})e^{ic^{-}_{p}u_{4}+ic^{-}_{q}u_{5}+ic^{-}_{r}u_{6}+iak_{1}(u_{4}-u_{5})+ibk_{2}(u_{5}-u_{6})+ick_{3}(u_{6}-u_{4})} \propto \nonumber \\
    \propto \int_{T_{0}}^{T} du_{4}\int_{}^{u_{4}}du_{5} \int_{}^{u_{4}}du_{6}e^{iu_{4}\big(c^{-}_{p}+ak_{1}-ck_{3}\big)+iu_{5}\big(c^{-}_{q}-ak_{1}+bk_{2}\big)+iu_{6}\big(c^{-}_{r}-bk_{2}+ck_{3}\big)} \propto \nonumber \\
    \propto \int_{T_{0}}^{T} du_{4}e^{iu_{4}\big(c^{-}_{p}+ak_{1}-ck_{3}+c^{-}_{q}-ak_{1}+bk_{2}+c^{-}_{r}-bk_{2}+ck_{3}\big)} \propto \delta \big(c^{-}_{p}+c^{-}_{q}+c^{-}_{r}\big) \ \ \
\end{eqnarray}
where we use that $\delta(K) \approx \int_{T_{0}}^{T}dxe^{iKx}$ since $T\rightarrow +\infty$ and $T_{0}\rightarrow -\infty$. Analogous structure will also have other terms in \eqref{TP:2} and \eqref{TP:3}. This brief analysis proofs our claim that
\begin{equation}
    \label{TP:6}
    n^{(i)}_{pqr} \propto \lambda^{3}\delta(\pm c^{-}_{p}\pm c^{-}_{q}\pm c^{-}_{r}).
\end{equation}
It is worth mentioning that due to the causal structure of Schwinger-Dyson diagrammatic technique we will have only linear secular growth of three point functions which comes from the aforementioned delta functions. It means that the corrections will be proportional to $\lambda^{3}(T-T_{0})$ and, hence, are suppressed by the extra power of $\lambda$  in comparison with the corrections to the two-point functions.

The same analysis can be used to estimate the growth of $n$--point functions. All of them have corrections proportional to the $\lambda^{n}(T-T_{0})$ and are suppressed as well. In concluding section we will clarify the origin of such a growth in loop corrections to the $n$--point functions.

\section{Loop corrections to the fermion propagators} \label{Section4}

\begin{figure}
    \centering
    \includegraphics[width=7.5cm]{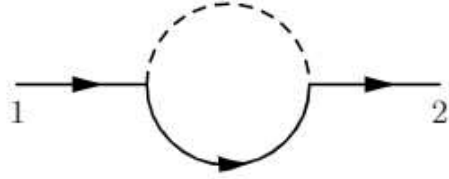}
    \caption{One loop corrections to the fermion Keldysh propagator. Solid lines correspond to the tree-level fermion propagators, dashed lines correspond to the tree--level boson propagators.}
    \label{fig3}
\end{figure}

In this section we show that one-loop corrections to the fermion Keldysh propagator (fig. \ref{fig3}) and to the vertexes do not possess secular growth. This supports the general observation that was made in \cite{46}.

\subsection{One loop correction to the Keldysh propagator for fermions}

In this subsection let us calculate the first loop correction to the Keldysh propagator for fermions\footnote{Again, it can be similarly shown that the loop corrections to the retarded and advanced propagators do
not grow in the limit $\dfrac{t_{1}+t_{2}}{2} \gg |t_{1}-t_{2}|$ \cite{16},\cite{2}.} which is given by the following expression:

\begin{eqnarray}
    \label{KGFF1}
    \Delta_{2} G^{K}(t_{1},x_{1};t_{2},x_{2})=\dfrac{\lambda^{2}}{2}\int d^{2}x_{3}d^{2}x_{4} \times \nonumber \\
     \times \bigg[
     G_{0}^{R}(t_{1},x_{1};t_{3},x_{3})G_{0}^{R}(t_{3},x_{3};t_{4},x_{4})D_{0}^{K}(t_{4},x_{4};t_{3},x_{3})G_{0}^{K}(t_{4},x_{4};t_{2},x_{2})+ \nonumber \\
      +G_{0}^{R}(t_{1},x_{1};t_{3},x_{3})G_{0}^{K}(t_{3},x_{3};t_{4},x_{4})D_{0}^{K}(t_{4},x_{4};t_{3},x_{3})G_{0}^{A}(t_{4},x_{4};t_{2},x_{2})+ \nonumber \\
      +G_{0}^{R}(t_{1},x_{1};t_{3},x_{3})G_{0}^{R}(t_{3},x_{3};t_{4},x_{4})D_{0}^{A}(t_{4},x_{4};t_{3},x_{3})G_{0}^{A}(t_{4},x_{4};t_{2},x_{2})+ \nonumber \\
      +G_{0}^{R}(t_{1},x_{1};t_{3},x_{3})G_{0}^{K}(t_{3},x_{3};t_{4},x_{4})D_{0}^{A}(t_{4},x_{4};t_{3},x_{3})G_{0}^{K}(t_{4},x_{4};t_{2},x_{2})+ \nonumber \\
      +G_{0}^{K}(t_{1},x_{1};t_{3},x_{3})G_{0}^{A}(t_{3},x_{3};t_{4},x_{4})D_{0}^{K}(t_{4},x_{4};t_{3},x_{3})G_{0}^{A}(t_{4},x_{4};t_{2},x_{2})+ \nonumber \\ +G_{0}^{R}(t_{1},x_{1};t_{3},x_{3})G_{0}^{A}(t_{3},x_{3};t_{4},x_{4})D_{0}^{R}(t_{4},x_{4};t_{3},x_{3})G_{0}^{A}(t_{4},x_{4};t_{2},x_{2})+ \nonumber \\
      +G_{0}^{K}(t_{1},x_{1};t_{3},x_{3})G_{0}^{K}(t_{3},x_{3};t_{4},x_{4})D_{0}^{R}(t_{4},x_{4};t_{3},x_{3})G_{0}^{A}(t_{4},x_{4};t_{2},x_{2})\bigg].
\end{eqnarray}

This expression follows from the diagrams of the form, which is depicted on the fig. \ref{fig3}.
Since the exact Keldysh propagator for fermions can be represented as follows
\begin{eqnarray}
    \label{KGFF2}
    G^{K}(1,2)=-i\int\dfrac{dp}{2\pi}\int\dfrac{dq}{2\pi}\bigg[\big(2\pi \delta(p-q)-2n^{\prime}_{qp}\big)\chi_{p}(1)\bar{\chi}_{q}(2)+\big(2\tilde{n}_{pq}-2\pi \delta(p-q)\big)\zeta_{p}(1)\bar{\zeta}_{q}(2)+\nonumber \\+
    2\kappa^{\prime}_{pq}\chi_{p}(1)\bar{\zeta}_{q}(2)+2\kappa^{\prime \dagger}_{pq}\zeta_{p}(1)\bar{\chi}_{q}(2)\bigg],
\end{eqnarray}
we will concentrate on the study of $n^{\prime}_{qp}$, $\tilde{n}_{pq}$ and $\tilde{\kappa}_{pq}$ in the limit $\dfrac{t_{1}+t_{2}}{2} \gg |t_{1}-t_{2}|$. They have the form:
\begin{eqnarray}
    \label{KGFF3}
    -2n^{\prime}_{qp}=i\dfrac{\lambda^{2}}{2}\int_{-\infty}^{+\infty}\dfrac{dq_{1}}{2\pi}\dfrac{1}{2|q_{1}|}\int_{0}^{+\infty}\dfrac{dk_{1}}{2\pi}  \int d^{2}x_{3}\int d^{2}x_{4} \times \nonumber \\
    \times \bigg[F(p,q,v,k_{1},q_{1})e^{ia(v_{3},0)/p+ia(v_{3},v_{4})/k_{1}-ia(v_{4},0)/q-ic^{+}_{q_{1}}(v_{4}-v_{3})}\times \nonumber \\
    \times\bigg(\theta(t_{1}-t_{3})\theta(t_{3}-t_{4})+\theta(t_{2}-t_{4})\theta(t_{4}-t_{3})+\theta(t_{1}-t_{3})\theta(t_{2}-t_{4})\bigg)e^{iu_{3}(p+k_{1}+c^{-}_{q_{1}})-iu_{4}(q+k_{1}+c^{-}_{q_{1}})}+ \nonumber \\
    +F(-p,-q,v,k_{1},q_{1})e^{ia(v_{3},0)/p-ia(v_{3},v_{4})/k_{1}-ia(v_{4},0)/q+ic^{+}_{q_{1}}(v_{4}-v_{3})}\times \nonumber \\
    \times\bigg(\theta(t_{1}-t_{3})\theta(t_{3}-t_{4})+\theta(t_{2}-t_{4})\theta(t_{4}-t_{3})-\theta(t_{1}-t_{3})\theta(t_{2}-t_{4})\bigg)e^{iu_{3}(p-k_{1}-c^{-}_{q_{1}})-iu_{4}(q-k_{1}-c^{-}_{q_{1}})}\bigg] \ \ , \ \
\end{eqnarray}
and
\begin{eqnarray}
    \label{KGFF4}
    2\tilde{n}_{pq}=-i\dfrac{\lambda^{2}}{2}\int_{-\infty}^{+\infty}\dfrac{dq_{1}}{2\pi}\dfrac{1}{2|q_{1}|}\int_{0}^{+\infty}\dfrac{dk_{1}}{2\pi}  \int d^{2}x_{3}\int d^{2}x_{4} \times \nonumber \\
    \times \bigg[F(p,q,v,k_{1},q_{1})e^{-ia(v_{3},0)/p-ia(v_{3},v_{4})/k_{1}+ia(v_{4},0)/q+ic^{+}_{q_{1}}(v_{4}-v_{3})}\times \nonumber \\
    \times\bigg(\theta(t_{1}-t_{3})\theta(t_{3}-t_{4})+\theta(t_{2}-t_{4})\theta(t_{4}-t_{3})+\theta(t_{1}-t_{3})\theta(t_{2}-t_{4})\bigg)e^{-iu_{3}(p+k_{1}+c^{-}_{q_{1}})+iu_{4}(q+k_{1}+c^{-}_{q_{1}})}+ \nonumber \\
    +F(-p,-q,v,k_{1},q_{1})e^{-ia(v_{3},0)/p+ia(v_{3},v_{4})/k_{1}+ia(v_{4},0)/q-ic^{+}_{q_{1}}(v_{4}-v_{3})}\times \nonumber \\
    \times\bigg(\theta(t_{1}-t_{3})\theta(t_{3}-t_{4})+\theta(t_{2}-t_{4})\theta(t_{4}-t_{3})-\theta(t_{1}-t_{3})\theta(t_{2}-t_{4})\bigg)e^{-iu_{3}(p-k_{1}-c^{-}_{q_{1}})+iu_{4}(q-k_{1}-c^{-}_{q_{1}})}\bigg] \ \ , \ \
\end{eqnarray}
and
\begin{eqnarray}
    \label{KGFF5}
    2\kappa^{\prime}_{pq}=i\dfrac{\lambda^{2}}{2}\int_{-\infty}^{+\infty}\dfrac{dq_{1}}{2\pi}\dfrac{1}{2|q_{1}|}\int_{0}^{+\infty}\dfrac{dk_{1}}{2\pi}  \int d^{2}x_{3}\int d^{2}x_{4} \times \nonumber \\
    \times \bigg[F(p,-q,v,k_{1},q_{1})e^{ia(v_{3},0)/p+ia(v_{3},v_{4})/k_{1}+ia(v_{4},0)/q-ic^{+}_{q_{1}}(v_{4}-v_{3})}\times \nonumber \\
    \times\bigg(-\theta(t_{1}-t_{3})\theta(t_{3}-t_{4})+\theta(t_{2}-t_{4})\theta(t_{4}-t_{3})-\theta(t_{1}-t_{3})\theta(t_{2}-t_{4})\bigg)e^{iu_{3}(p+k_{1}+c^{-}_{q_{1}})+iu_{4}(-q+k_{1}+c^{-}_{q_{1}})}+ \nonumber \\
    +F(-p,q,v,k_{1},q_{1})e^{ia(v_{3},0)/p-ia(v_{3},v_{4})/k_{1}+ia(v_{4},0)/q+ic^{+}_{q_{1}}(v_{4}-v_{3})}\times \nonumber \\
    \times\bigg(-\theta(t_{1}-t_{3})\theta(t_{3}-t_{4})+\theta(t_{2}-t_{4})\theta(t_{4}-t_{3})-\theta(t_{1}-t_{3})\theta(t_{2}-t_{4})\bigg)e^{iu_{3}(p-k_{1}-c^{-}_{q_{1}})-iu_{4}(-q-k_{1}-c^{-}_{q_{1}})}\bigg], \ \
\end{eqnarray}
where
\begin{equation}
    \label{KGFF6}
    F(p,q,v,k_{1},q_{1})=\dfrac{\Phi(v_{3})\Phi(v_{4})}{2}\bigg(\dfrac{1}{k_{1}}-\dfrac{1}{p}\bigg)\bigg(\dfrac{1}{k_{1}}-\dfrac{1}{q}\bigg),
\end{equation}
and unlike the case of scalar counterparts in these expressions we drop the upper index to simplify them. Integrals in \eqref{KGFF3} -- \eqref{KGFF5} are very similar to those calculated in Appendix \ref{A}, so we show here only the results
\begin{equation}
    \label{KGFF7}
    \begin{cases}
n^{\prime}_{qp} \propto \delta(p-q)\delta(p+k_{1}+c^{-}_{q_{1}})\sim 0 \\
\tilde{n}_{pq} \propto \delta(p-q)\delta(p+k_{1}+c^{-}_{q_{1}})\sim 0 \\
\kappa^{\prime}_{pq} \propto \delta(p+q) \sim 0
    \end{cases} \ , \qquad \text{in the limit} \qquad \dfrac{t_{1}+t_{2}}{2}=T\rightarrow +\infty,
\end{equation}
since $p,q,k_{1}$ and $c^{-}_{q_{1}}$ take their values in the interval $(0,+\infty)$. Thus, the Keldysh propagator for fermions does not receive secularly growing corrections, which agrees with the observations made in \cite{46}.

Now, it is worth to make the following observation. We see that in the theory under consideration the secular growth comes only from such diagrams which contain only external scalar legs. Indeed, such legs contain delta function of the form $\delta\big(c^{-}_{p} \pm c^{-}_{q} \pm ... \big)$ where $p,q,...$ are momenta of these legs. In its turn the argument of such a delta function can vanish for all positive values of $p,q,...$ which leads to the multiplier of the form $\delta(0)=2\pi(T-T_{0})$. This property disappears when there are at least two fermion external legs. This is because  the delta functions acquire the form $\delta(k\pm k^{\prime} \pm...)$ where   $k,k^{\prime}, ...$ are momenta of external fermions (which take only positive values as can be seen from the mode expansion \eqref{eq:i2}: pay attention to the limits of integration). The argument of the latter delta function cannot be equal to zero for an entire range of values momenta.

\subsection{One loop corrections to vertexes}\label{5}

\begin{figure}
    \centering
    \includegraphics[width=7.5cm]{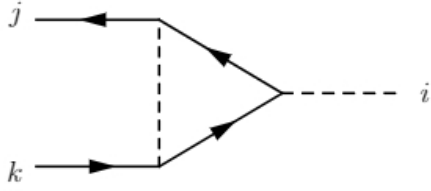}
    \caption{One loop corrections to vertexes.}
    \label{fig4}
\end{figure}

In this subsection we show that one-loop corrections to the vertexes depicted on the fig. \ref{fig4} do not possess growing  with
time contributions in the limit $t_{1} \approx t_{2} \approx t_{3} \approx T\rightarrow +\infty$. The calculation of the corrections to the vertexes is very similar to the one in the section   \ref{3.3}. The difference is that now we have one scalar leg and two fermion legs. In analogy with \eqref{TP:2} we represent correlation functions as
\begin{eqnarray}
    \label{EQ:2}
  \langle\phi_{i}(t_{1},x_{1})\psi_{j}(t_{2},x_{2})\psi_{k}(t_{3},x_{3}) \rangle=\int_{-\infty}^{+\infty}\dfrac{dp}{2\pi}\int_{0}^{+\infty}\dfrac{dk}{2\pi}\dfrac{dk^{\prime}}{2\pi}\bigg[N^{(1)}_{ijk}(p,k,k^{\prime})f_{p}(1)\chi_{k}(2)\chi_{k^{\prime}}(3)+\nonumber \\
  +N^{(2)}_{ijk}(p,k,k^{\prime})f^{*}_{p}(1)\chi_{k}(2)\chi_{k^{\prime}}(3)
    +N^{(3)}_{ijk}(p,k,k^{\prime})f_{p}(1)\zeta_{k}(2)\chi_{k^{\prime}}(3)+...\bigg],
\end{eqnarray}
and calculate one loop corections to the $N^{(s)}_{ijk}(p,k,k^{\prime})$ with $s=1,...,8$ and $i,j,k=1,2$. The result of calculations has the form
\begin{equation}
    \label{EQ:1}
    N^{(s)}_{ijk}(p,k,k^{\prime}) \propto \lambda^{3} \delta(\pm k \pm k^{\prime} \pm c^{-}_{p}),
\end{equation}
where the signs of $k,k^{\prime}$ and $c^{-}_{p}$ depends on the values of multi-index $s;i,j,k$. As was discussed above, the argument of such a delta function can not be equal to zero on an entire interval of values of external momenta: the argument of this delta function can vanish only for some singular values of $k,k^{\prime}$ and $q$. Since the growth in $T$ can come only from delta function, whose argument is vanishing on the entire interval of values of momenta, it means that there is no secular growth of the loop corrections to the vertexes.



\section{Dyson equation for the exact Keldysh propagator for scalars}\label{Section5}

In this section we sum up the leading contributions from all loops to the anomalous quantum average of the scalars, $\kappa_{pq}$, with the help of Dyson-Schwinger equations. Also, we find the stationary solution of this equation and its contribution to the exact scalar Keldysh propagator.

Let us briefly remind the main results of the previous sections. We found that the anomalous average $\kappa^{(2)}_{pq}$ in the Keldysh propagator for scalars receives secularly growing loop corrections proportional to $\lambda^{2}(T-T_{0})$, while other quantities  $n^{(2)}_{qp}$, $n^{\prime}_{pq}$, $\tilde{n}_{pq}$ and $\kappa^{\prime}_{pq}$ both for fermions and scalars do not receive such corrections. Since for big enough value of $T-T_{0}$ the product $\lambda^{2}(T-T_{0})$ becomes large even for small $\lambda$, loop corrections are not suppressed in comparison with the tree-level contributions.

That is the reason why to understand the physics behind such a growth one at least has to sum up all the terms which are powers of $\lambda^{2}(T-T_{0})$ and to drop terms which are suppressed by higher powers of $\lambda$, i.e. such as $\lambda^{3}(T-T_{0})$ or $\lambda\big(\lambda^{2}(T-T_{0})\big)^{2}$ and etc..

In order to do that, in principle one has to solve the system of Dyson-Schwinger equations for the exact
propagators, $D^{K},D^{R},D^{A}$ and $G^{K},G^{R},G^{A}$ and for the vertexes in the limit $T=\dfrac{t_{1}+t_{2}}{2} \gg |t_{1}-t_{2}|$ (where $t_{1,2}$ are the arguments of $D^{K(R,A)}$, $G^{K(R,A)}$), i.e. as $T-T_{0} \rightarrow +\infty$. But taking into account that all vertexes, retarded, advanced propagators and also the Keldysh
propagator for the fermions receive subleading corrections in powers of $\lambda$, we can put them to their tree–level
values in the system of Dyson–Schwinger equations.

Thus, if we are interested only in the leading
corrections, this system reduces to the single equation for the exact Keldysh propagator for bosons or even, more concretely, for its part represented by $\kappa_{pq}$ with positive $p$ and $q$. Then, to obtain the equation one simply can make the change $D_{0}^{K}(4,2) \rightarrow D^{K}(4,2)$ in \eqref{eq:22}:
\begin{eqnarray}
\label{eq:105}
     D^{K}(t_{1},x_{1};t_{2},x_{2})=D_{0}^{K}(t_{1},x_{1};t_{2},x_{2})+\dfrac{\lambda^{2}}{2}\int d^{2}x_{3}d^{2}x_{4} \times \nonumber \\
       \times \bigg[
     D_{0}^{K}(t_{1},x_{1};t_{3},x_{3})G_{0}^{K}(t_{3},x_{3};t_{4},x_{4})G_{0}^{R}(t_{4},x_{4};t_{3},x_{3})D_{0}^{A}(t_{4},x_{4};t_{2},x_{2})+ \nonumber \\
      +D_{0}^{K}(t_{1},x_{1};t_{3},x_{3})G_{0}^{A}(t_{3},x_{3};t_{4},x_{4})G_{0}^{K}(t_{4},x_{4};t_{3},x_{3})D_{0}^{A}(t_{4},x_{4};t_{2},x_{2})+ \nonumber \\
      +D_{0}^{R}(t_{1},x_{1};t_{3},x_{3})G_{0}^{R}(t_{3},x_{3};t_{4},x_{4})G_{0}^{K}(t_{4},x_{4};t_{3},x_{3})D^{K}(t_{4},x_{4};t_{2},x_{2})+ \nonumber \\
       +D_{0}^{R}(t_{1},x_{1};t_{3},x_{3})G_{0}^{K}(t_{3},x_{3};t_{4},x_{4})G_{0}^{A}(t_{4},x_{4};t_{3},x_{3})D^{K}(t_{4},x_{4};t_{2},x_{2})+ \nonumber \\
      +D_{0}^{R}(t_{1},x_{1};t_{3},x_{3})G_{0}^{R}(t_{3},x_{3};t_{4},x_{4})G_{0}^{A}(t_{4},x_{4};t_{3},x_{3})D_{0}^{A}(t_{4},x_{4};t_{2},x_{2})+ \nonumber \\
 +D_{0}^{R}(t_{1},x_{1};t_{3},x_{3})G_{0}^{K}(t_{3},x_{3};t_{4},x_{4})G_{0}^{K}(t_{4},x_{4};t_{3},x_{3})D_{0}^{A}(t_{4},x_{4};t_{2},x_{2})+ \nonumber \\
+D_{0}^{R}(t_{1},x_{1};t_{3},x_{3})G_{0}^{A}(t_{3},x_{3};t_{4},x_{4})G_{0}^{R}(t_{4},x_{4};t_{3},x_{3})D_{0}^{A}(t_{4},x_{4};t_{2},x_{2})\bigg],
 \end{eqnarray}
where the ansatz for the exact propagator is
\begin{eqnarray}
    \label{eq:106}
    D^{K}(t_{1},x_{1};t_{2},x_{2}) = D_{0}^{K}(t_{1},x_{2};t_{2},x_{2})+(-i) \int_{-\infty}^{+\infty} \dfrac{dp}{2\pi}\int_{-\infty}^{+\infty} \dfrac{dq}{2\pi} \bigg[ 2\kappa_{pq}(t_{1},t_{2})f_{p}(1)f_{q}(2) + h.c.\bigg],
\end{eqnarray}
and we neglect the term $n_{qp}$ since it is supressed in comparison with $\kappa_{pq}$. Then the equation reduces to
\begin{eqnarray}
    \label{eQ:107}
    2\kappa_{pq}(T)=2\kappa^{(2)}_{pq}(T)+ \nonumber \\
    +\dfrac{1}{2}\bigg[(-i)\int_{-\infty}^{+\infty}\dfrac{dq_{1}}{2\pi}\int d^{2}x_{3}\int d^{2}x_{4}\bigg\{2\kappa_{q_{1}q}(t_{4},t_{2})\theta(t_{1}-t_{3})f^{*}_{p}(3)f_{q_{1}}(4)G_{0}^{R}(3,4)G_{0}^{K}(4,3)+\nonumber \\
    +2\kappa_{q_{1}q}(t_{4},t_{2})\theta(t_{1}-t_{3})f^{*}_{p}(3)f_{q_{1}}(4)G_{0}^{K}(3,4)G_{0}^{A}(4,3)\bigg\} + (p \leftrightarrow q) \bigg],
\end{eqnarray}
where $\kappa^{(2)}_{pq}(t_{1},t_{2})$ is defined in \eqref{17} and $(p \leftrightarrow q)$ means the same expression as explicitly written with the exchange of $p$ and $q$. Note that $\kappa_{pq}$ is symmetric under the exchange of $p$ and $q$ and the obtained equation respects this symmetry.

Furthermore, we assume that $\kappa_{pq}(T)$ is a slow function in comparison with modes (this is the kinetic approximation). That is why we do the following simplification
\begin{equation}
    \label{equation1}
    \begin{cases}
    \kappa_{pq}(t_{1},t_{2}) \approx \kappa_{pq}\bigg(\dfrac{t_{1}+t_{2}}{2}\bigg) \approx \kappa_{pq}(T)\\
     \kappa_{q_1 q}(t_{4},t_{2}) \approx \kappa_{q_1 q} \bigg(\dfrac{t_{4}+t_{2}}{2}\bigg) \approx \kappa_{q_1 q}(T)\\
    \end{cases}.
\end{equation}
In Appendix \ref{B} it is shown that Dyson-Schwinger equation for $\kappa_{pq}$ reduces to the following form:
\begin{equation}
    \label{eQ:126}
    \partial_{T}\kappa_{pq}(T)=F(p,q)+\dfrac{1}{2}\bigg[\int_{0}^{+\infty}\dfrac{dq_{1}}{2\pi}F(p,-q_{1})\kappa_{q_{1}q}(T)+(p \leftrightarrow q)\bigg], \qquad p,q\geq0.
\end{equation}
where $F(p,q)$ is defined in eq. \eqref{eq:tpp4}.

\subsection{Stationary solution of the Dyson-Schwinger equation}

Let us find the stationary solution of the equation \eqref{eQ:126}:
\begin{equation}
    \label{eQ:127}
    F(p,q)+\dfrac{1}{2}\bigg[\int_{0}^{+\infty}\dfrac{dq_{1}}{2\pi}F(p,-q_{1})\kappa_{q_{1}q}+(p \leftrightarrow q)\bigg]=0
\end{equation}
Formally, it looks like
\begin{equation}
    \label{eq:127}
    \kappa_{pq}=-2\pi\delta(p+q), \quad p, q \geq 0.
\end{equation}
Substituting this expression into the exact propagator \eqref{eq:106}, we find :
\begin{equation}
    \label{eq:128}
    D^{K}(t_{1},x_{1};t_{2},x_{2})=D_{0}^{K}(t_{1},x_{1};t_{2},x_{2}) + \int_{0}^{+\infty}\dfrac{dp}{2\pi}\dfrac{dq}{2\pi}2\kappa_{pq}\dfrac{1}{\sqrt{2|p|}}e^{-i|p|t_{1}+ipx_{1}}\dfrac{1}{\sqrt{2|q|}}e^{-i|q|t_{2}+iqx_{2}}+ c.c.
\end{equation}
Since the static solution $\kappa_{pq}$ is not zero only when $p=q=0$ we have to carefully calculate the last expression:

\begin{eqnarray}
    \label{eQ:EPK1}
    \int_{0}^{+\infty}\dfrac{dp}{2\pi}\dfrac{dq}{2\pi}2\kappa_{pq}\dfrac{1}{\sqrt{2|p|2|q|}}e^{-i|p|t_{1}+ipx_{1}}e^{-i|q|t_{2}+iqx_{2}}+c.c.=\nonumber \\
    =\int_{0}^{+\infty}\dfrac{dp}{2\pi}\dfrac{dq}{2\pi}\big(-2\pi \delta(p+q)\big)\dfrac{1}{\sqrt{|pq|}}e^{-i|p|t_{1}+ipx_{1}}e^{-i|q|t_{2}+iqx_{2}}+c.c. = \nonumber \\
        =-\dfrac{1}{2\pi}\int_{0}^{+\infty}dy_{+}\int_{-\infty}^{+\infty}dy_{-}\dfrac{\delta(y_{+})}{\sqrt{|y^{2}_{+}-y^{2}_{-}|}}e^{-i|y_{-}|T+iy_{-}X}+c.c.
\end{eqnarray}
where
\begin{equation}
    \label{666}
    \begin{cases}
    T=\dfrac{1}{\sqrt{2}}\big(t_{1}+t_{2}\big) \\
    X=\dfrac{1}{\sqrt{2}}\big(x_{1}-x_{2}\big),
    \end{cases}
\end{equation}
and we have made the following change of the integration variables: $y_{\pm}=\dfrac{p\pm q}{\sqrt{2}}$.

Using that $\int_{0}^{+\infty}dy_{+}\delta(y_{+})=\dfrac{1}{2}$ we get
\begin{eqnarray}
     \label{6666}
     -\dfrac{1}{2\pi}\int_{0}^{+\infty}dy_{+}\int_{-\infty}^{+\infty}dy_{-}\dfrac{\delta(y_{+})}{\sqrt{|y^{2}_{+}-y^{2}_{-}|}}e^{-i|y_{-}|T+iy_{-}X}+c.c.=-\dfrac{1}{2\pi}\int_{-\infty}^{+\infty}dy_{-}\dfrac{\cos{\big(|y_{-}|T-yX\big)}}{\sqrt{y^{2}_{-}}}=\nonumber \\
     =-\dfrac{1}{2\pi}\int_{0}^{+\infty}dy_{-}\dfrac{\cos{(y_{-}(T-X))}}{y_{-}}-\dfrac{1}{2\pi}\int_{0}^{+\infty}dy_{-}\dfrac{\cos{(y_{-}(T+X))}}{y_{-}}=\nonumber\\
     =-\dfrac{1}{2\pi}\lim_{\epsilon \to 0}\int_{\epsilon|T-X|}^{+\infty}dy\dfrac{\cos{y}}{y}-\dfrac{1}{2\pi}\lim_{\epsilon \to 0}\int_{\epsilon|T+X|}^{+\infty}dy\dfrac{\cos{y}}{y}= \nonumber\\
     =\dfrac{1}{2\pi}\lim_{\epsilon \to 0}ci(\epsilon|T-X|)+\dfrac{1}{2\pi}\lim_{\epsilon \to 0}ci(\epsilon|T+X|), \ \ \
\end{eqnarray}
where $ci(z)\equiv -\int_{z}^{+\infty}dy\dfrac{\cos{y}}{y}$ is the cosine integral. Using asymptotic expansion of the cosine integral \cite{101}:
\begin{equation}
    \lim_{z \to 0}ci(z)\approx \ln{z}
\end{equation}
we get that
\begin{eqnarray}
     \label{66666}
     \int_{0}^{+\infty}\dfrac{dp}{2\pi}\dfrac{dq}{2\pi}2\kappa_{pq}\dfrac{1}{\sqrt{2|p|2|q|}}e^{-i|p|t_{1}+ipx_{1}}e^{-i|q|t_{2}+iqx_{2}}\approx \dfrac{1}{\pi}\lim_{\epsilon \to 0} \ln{\epsilon T}
\end{eqnarray}
in the limit $T \equiv (t_1+t_2)/\sqrt{2} \to \infty$. In all, we obtain that the resummed Keldysh propagator acquires the following form:

\begin{equation}
    \label{eq:EPK4}
    D^{K}(t_{1},x_{1};t_{2},x_{2}) \approx D_{0}^{K}(t_{1}-t_{2},x_{1} - x_{2}) + \dfrac{1}{\pi}\ln{\epsilon \, (t_1+t_2)},
\end{equation}
if only the leading contributions are kept in the limit in question.

Thus, the exact propagator receives a shift, where the divergence is similar to the one appearing in the massless two--dimensional scalar propagator. Here we see an explicit breaking of the Poincar\'e invariance due to the presence of non--zero $\kappa_{pq}$, which is appearing due to the background scalar field.

It is tempting to interpret this divergent constant as due to the presence of the condensate, which appears because the minimum of the effective potential $V_{eff}(\phi)$ leading to the \eqref{eq:opp3} is not at $\langle \phi \rangle =0$. \textcolor{black}{In fact, in the initial state we have that the anomalous quantum average $\kappa_{pq}$ is zero, because in such a state $\hat{\alpha}_p |0\rangle = 0$, by definition. But, once $\kappa_{pq} \equiv \langle \hat{\alpha}_p \, \hat{\alpha}_q \rangle \neq 0$ that already means a presence of a condensate.} Furthermore, the standard way to define the condensate is to consider the following limit of the two-point function: $\lim_{|x_{1}-x_{2}| \to \infty}D(x_{1}-x_{2})$. If this limit is a non--zero constant there is a condensate\footnote{These observations are also supported by the considerations of the section 3.1.}. That is all simple and true if the propagator decays with the distance, as is the case in dimensions higher than two. In two dimensions, however, $D^{K}_{0}(\underline{x}_{1},\underline{x}_{2}) \propto \ln{|\underline{x}_{1}-\underline{x}_{2}|}$ and does not decay with the distance. This fact complicates our study. Nevertheless, we still think that our calculations here reveal the presence of the condensate in the theory, as we clarify in the concluding section below. \textcolor{black}{Also in appendix \ref{3D} we extend our considerations to the three dimensional case to see what does the situation under consideration mean in higher dimensions. There we also encounter an expectation value that depends on the average time $T$.}

\section{Conclusions}\label{Section6}

First, we calculated one loop corrections to correlation functions for scalars with the use of the Keldysh-Schwinger technique and concentrated our analysis on secularly growing contributions. We showed that calculation of the one loop correction to the one point scalar correlation function is consistent with the obtained in \cite{46} effective potential for scalars.

Second, we have found one loop correction to the Keldysh propagator for scalars and showed that the anomalous expectation value $\kappa_{pq} \equiv \langle\hat{\alpha}_{p}\hat{\alpha}_{q} \rangle $ receives secularly growing contribution proportional to $\lambda^{2}(T-T_{0})$, only if both $p,q$ are positive. I.e. we find a growth of the anomalous average only for those modes which are propagating along the background field $\phi(t-x)$. At the same time the expectation value of the level population $n_{pq}\equiv \langle\hat{\alpha}^{\dagger}_{p}\hat{\alpha}_{q} \rangle $ does not receive growing with time corrections at all. Also, we calculated one loop corrections to the three-point scalar correlation functions and found that they also receives growing with time contributions. The latter, however, are suppressed by the higher power of  $\lambda$, i.e. they are proportional to the $\lambda^{3}(T-T_{0})$.

Third, we repeated the same calculations for the correlation functions of fermions and found that neither Keldysh propagator nor vertexes receive secularly growing loop corrections. This supports the general observation that was made in \cite{46}.

After that we derived integro-differential Dyson--Schwinger equation for $\kappa_{pq}$, which is not trivial only for positive values of $p$ and $q$. This equation sums up secularly growing contributions from all loops, i.e. --- corrections of the form $\big(\lambda^{2}(T-T_{0})\big)^{n}$, where $n$ is the number of loops. Even though, a general solution of Dyson--Schwinger equation was not found even in the kinetic approximation, still we have obtained the stationary form of $\kappa_{pq}$. Its contribution to the exact Keldysh propagator gives a contribution, which we interpret as the scalar condensate. The latter appears due to the non--trivial minimum of the effective potential, which in turn is generated due to the quantum fluctuations of fermions.

Let us make here some observations, which may shed some light on the behaviour of the loop corrections to the correlation functions. As was mentioned several times, in \cite{46} the effective potential for scalars was obtained for large and slowly changing background field $\Phi$:

\begin{equation}
    \label{C1}
    S_{eff} \approx \int d^{2}x \bigg[\dfrac{1}{2}\big(\partial\varphi\big)^{2} - V_{eff}(\varphi)\bigg], \qquad \text{where} \qquad V_{eff}(\lambda \varphi) \simeq \dfrac{(\lambda \varphi)^{2}}{2\pi}\ln\dfrac{\lambda \varphi}{\Lambda}-\dfrac{(\lambda \varphi)^{2}}{4\pi}.
\end{equation}
The scalar field $\varphi$ in this effective theory can be represented as follows:

\begin{equation}
    \label{C2}
    \lambda\varphi(t,x)=\Phi(v)+\lambda\hat{\phi}(t,x),
\end{equation}
where $\Phi(v)$ is the classical background field under consideration and $\hat{\phi}$ represents quantum fluctuations. Then, we expand in series the effective action around the classical field $\Phi(v)$:
\begin{equation}
    \label{C3}
    \Delta \hat{S}_{eff} \sim \int d^{2}x V_{eff}(\Phi)+\lambda\int d^{2}x V^{\prime}_{eff}(\Phi)\hat{\phi}+\lambda^{2}\int d^{2}x \dfrac{1}{2!}V^{\prime \prime}_{eff}(\Phi)\hat{\phi}\hat{\phi}+\lambda^{3}\int d^{2}x \dfrac{1}{3!}V^{\prime \prime \prime}_{eff}(\Phi)\hat{\phi}\hat{\phi}\hat{\phi}+...
\end{equation}
Using the exact form of $\hat{\phi}$ \eqref{eq:11} we get that the first term in \eqref{C3} has the form
\begin{eqnarray}
     \label{C4}
    \lambda\int d^{2}x V^{\prime}_{eff}(\Phi)\hat{\phi}=\lambda\int_{-\infty}^{+\infty}\dfrac{dp}{2\pi}\int du\int dv V^{\prime}_{eff}(\Phi(v))\bigg(\dfrac{1}{\sqrt{2|p|}}e^{-ic^{-}_{p}u-ic^{+}_{p}v}\hat{\alpha}_{p}+h.c.\bigg)=\nonumber \\
    =\lambda \int_{-\infty}^{+\infty}\dfrac{dp}{2\pi}\dfrac{2\pi \delta(c^{-}_{p})}{\sqrt{2|p|}}\bigg\{V_{1}(p)\hat{\alpha}_{p}+h.c.\bigg\},
\end{eqnarray}
where
\begin{equation}
\label{mopou5t}
    V_{n}(p) \equiv \dfrac{1}{n!}\int_{-\infty}^{+\infty}dv \, V^{(n)}_{eff}\big(\Phi(v)\big)e^{i\sqrt{2}pv}.
\end{equation}
Using that
\begin{equation}
    \label{eq0987654312}
    \begin{cases}
    \delta(c^{-}_{p})=\sqrt{2}\delta(|p|-p)=\sqrt{2}\delta(0) \qquad \qquad \  \text{if} \ \ \ \ p \geq 0\\
    \delta(c^{-}_{p})=\sqrt{2}\delta(|p|-p)=\sqrt{2}\delta(2p)=0 \qquad \text{if} \ \ \ \ p < 0
    \end{cases}
\end{equation}
we get
\begin{eqnarray}
     \label{123C4}
    \lambda\int d^{2}x V^{\prime}_{eff}(\Phi)\hat{\phi}=\lambda \int_{0}^{+\infty}\dfrac{dp}{2\pi}\dfrac{2\pi \delta(0)}{\sqrt{p}}\bigg\{V_{1}(p)\hat{\alpha}_{p}+h.c.\bigg\}.
\end{eqnarray}
The second term looks like
\begin{eqnarray}
      \label{C5}
   \lambda^{2} \int d^{2}x \dfrac{1}{2!}V^{\prime \prime}_{eff}(\Phi)\hat{\phi}\hat{\phi}    =\lambda^{2}\int_{0}^{+\infty}\dfrac{dp}{2\pi}\int_{0}^{+\infty}\dfrac{dq}{2\pi}\dfrac{2\pi \delta(0)}{\sqrt{4pq}}\bigg\{V_{2}(p+q)\hat{\alpha}_{p}\hat{\alpha}_{q}+V_{2}(p-q)\hat{\alpha}_{p}\hat{\alpha}^{\dagger}_{q}+h.c.\bigg\}+\nonumber \\
    +\lambda^{2}\int_{-\infty}^{0}\dfrac{dp}{2\pi}\dfrac{1}{\sqrt{2}}\dfrac{1}{\sqrt{4p^{2}}}\bigg\{V_{2}(0)\hat{\alpha}_{p}\hat{\alpha}^{\dagger}_{p}+h.c.\bigg\}, \ \ \
\end{eqnarray}
and the third term has the form
\begin{eqnarray}
     \label{C6}
     \lambda^{3}\int d^{2}x \dfrac{1}{3!}V^{\prime \prime \prime}_{eff}(\Phi)\hat{\phi}\hat{\phi}\hat{\phi}=\lambda^{3}\int_{0}^{+\infty}\dfrac{dp}{2\pi}\int_{0}^{+\infty}\dfrac{dq}{2\pi}\int_{0}^{+\infty}\dfrac{dr}{2\pi}\dfrac{2\pi \delta(0)}{\sqrt{8pqr}}\bigg\{V_{3}(p+q+r)\hat{\alpha}_{p}\hat{\alpha}_{q}\hat{\alpha}_{r}+h.c.\bigg\} +\nonumber \\
     +\lambda^{3}\int_{0}^{+\infty}\dfrac{dp}{2\pi}\int_{0}^{+\infty}\dfrac{dq}{2\pi}\int_{0}^{+\infty}\dfrac{dr}{2\pi}\dfrac{2\pi \delta(0)}{\sqrt{8pqr}}\bigg\{V_{3}(p+q-r)\big[\hat{\alpha}_{p}\hat{\alpha}_{q}\hat{\alpha}^{\dagger}_{r}+\hat{\alpha}_{p}\hat{\alpha}^{\dagger}_{r}\hat{\alpha}_{q}+\hat{\alpha}^{\dagger}_{r}\hat{\alpha}_{p}\hat{\alpha}_{q}\big]+h.c.\bigg\}+ \nonumber \\
     +\lambda^{3}\int_{0}^{+\infty}\dfrac{dp}{2\pi}\int_{-\infty}^{0}\dfrac{dq}{2\pi}\dfrac{\sqrt{2}}{\sqrt{8pq^{2}}}\bigg\{V_{3}(p)\big[\hat{\alpha}_{p}\hat{\alpha}_{q}\hat{\alpha}^{\dagger}_{q}+\hat{\alpha}^{\dagger}_{q}\hat{\alpha}_{q}\hat{\alpha}_{p}+\hat{\alpha}_{p}\hat{\alpha}^{\dagger}_{q}\hat{\alpha}_{q}\big]+h.c.\bigg\}+\nonumber \\
     +\lambda^{3}\int_{-\infty}^{0}\dfrac{dp}{2\pi}\dfrac{dq}{2\pi}\dfrac{V_{3}(0)}{\sqrt{2}}\dfrac{1}{\sqrt{8pq|p+q|}}\bigg\{\hat{\alpha}_{p}\hat{\alpha}_{q}\hat{\alpha}^{\dagger}_{p+q}+\hat{\alpha}^{\dagger}_{p+q}\hat{\alpha}_{q}\hat{\alpha}_{p}+\hat{\alpha}_{p}\hat{\alpha}^{\dagger}_{p+q}\hat{\alpha}_{q} +h.c.\bigg\}, \ \
\end{eqnarray}
The singular expression $\delta(0)$ appears after the integration over the $u$-coordinate\footnote{It is interesting to point out, that if one quantize scalars in light-cone coordinates, where, for example $v$ is time, then integrating over $u$ coordinate and droping the integration over $v$ coordinate would give the part of the Hamiltonian. However, that is not the case in the present situation since we use the canonical quantization with the time $t$.}. In fact, the modes for scalars in \eqref{eq:11} can be represented as $f_{p}=\dfrac{1}{\sqrt{2|p|}}e^{-iuc^{-}_{p}-ivc^{+}_{p}}$, where $c^{\pm}_{p}=\dfrac{1}{\sqrt{2}}(|p|\pm p)$. Then, the integration over the $u$-coordinate can give only the delta functions of the following form $\delta(c^{-}_{p} \pm c^{-}_{q}+...)$, which for positive values of momenta $p,q,...$ become $\delta(0)$. As we have discussed in the main body of the paper this is exactly the same expression which leads to the secularly growing factor $T-T_0$. The difference is that in the operator formalism in the Schwinger--Keldysh technique the limits of integration are always finite (i.e. $T$ and $T_0$), while in the Feynman effective action \eqref{C1} we as usual put them to infinity.

Since the structure of such terms is the same at any order of $\lambda$, we concentrate on the first few only. Writing terms which contain $\delta(0)$ as the multiplier and are of order $\lambda^{2}$ and $\lambda^{3}$ only, we get that

\begin{eqnarray}
     \label{C10}
     \hat{S}_{eff}=...+\lambda^{2}\int_{0}^{+\infty}\dfrac{dp}{2\pi}\int_{0}^{+\infty}\dfrac{dq}{2\pi}\dfrac{2\pi \delta(0)}{\sqrt{4pq}}\bigg\{V_{2}(p+q)\hat{\alpha}_{p}\hat{\alpha}_{q}+V_{2}(p-q)\hat{\alpha}_{p}\hat{\alpha}^{\dagger}_{q}+h.c.\bigg\}+ \nonumber \\
     +\lambda^{3}\int_{0}^{+\infty}\dfrac{dp}{2\pi}\int_{0}^{+\infty}\dfrac{dq}{2\pi}\int_{0}^{+\infty}\dfrac{dr}{2\pi}\dfrac{2\pi \delta(0)}{\sqrt{8pqr}}\bigg\{V_{3}(p+q+r)\hat{\alpha}_{p}\hat{\alpha}_{q}\hat{\alpha}_{r}+h.c.\bigg\}+\nonumber \\
     +\lambda^{3}\int_{0}^{+\infty}\dfrac{dp}{2\pi}\int_{0}^{+\infty}\dfrac{dq}{2\pi}\int_{0}^{+\infty}\dfrac{dr}{2\pi}\dfrac{2\pi \delta(0)}{\sqrt{8pqr}}\bigg\{V_{3}(p+q-r)\big[\hat{\alpha}_{p}\hat{\alpha}_{q}\hat{\alpha}^{\dagger}_{r}+\hat{\alpha}_{p}\hat{\alpha}^{\dagger}_{r}\hat{\alpha}_{q}+\hat{\alpha}^{\dagger}_{r}\hat{\alpha}_{p}\hat{\alpha}_{q}\big]+h.c.\bigg\}+...\ \ \
\end{eqnarray}
Analyzing the structure of \eqref{C4}-\eqref{C6} it is clear that the terms $\hat{\alpha}_{p}$, $\hat{\alpha}_{p}\hat{\alpha}_{q}$, $\hat{\alpha}_{p}\hat{\alpha}_{q}\hat{\alpha}_{r}$, etc. are present because the background field is $v$-dependent. At the same time in the constant ``background field'', $\Phi(v)=M=const$, the integration over $t$ and $x$ would lead to the delta--functions establishing energy and momentum conservation. This means that anomalous terms proportional to $\hat{\alpha}_{p}$, $\hat{\alpha}_p \hat{\alpha}_q$, $\hat{\alpha}_p \hat{\alpha}_q \hat{\alpha}_r$ and etc. are forbidden in such a situation. While in the non--trivial $\Phi(v)$ background they are allowed.


Let us calculate as an example the first few terms of the series expansion for the effective action, since there are some subtleties. In the situation under consideration, the integration over $u$ and  $v$ coordinates will give the following delta functions
\begin{equation}
    \label{e5t71i9}
    \delta\big(c^{-}_{p_{1}}+...+c^{-}_{p_{n}}\big)\delta \big(c^{+}_{p_{1}}+...+c^{+}_{p_{n}}\big)=\delta(|p_{1}|+...|p_{n}|)\delta(p_{1}+...+p_{n})
\end{equation}
which exactly reveals the conservation of energy and momentum.
At the same time, the term
\begin{eqnarray}
    \label{C12}
    \lambda\int d^{2}x V^{\prime}_{eff}(M)\hat{\phi}
\end{eqnarray}
contains a divergent multiplier. To get rid of this linear term one can choose the minimum of the effective potential, $V'_{eff}(M) = 0$, rather than an arbitrary constant $M$.

Next, consider the term of the form
\begin{eqnarray}
     \label{C138}
     \lambda^{2}\int_{0}^{+\infty}\dfrac{dp}{2\pi}\int_{0}^{+\infty}\dfrac{dq}{2\pi}\dfrac{2\pi \delta\big(c^{-}_{p}+c^{-}_{q}\big)}{\sqrt{4pq}}\bigg\{V_{2}(p+q)\hat{\alpha}_{p}\hat{\alpha}_{q}+h.c.\bigg\}=\nonumber \\
     = \lambda^{2}\int_{0}^{+\infty}\dfrac{dp}{2\pi}\int_{0}^{+\infty}\dfrac{dq}{2\pi}\dfrac{(2\pi)^{2} \delta\big(c^{-}_{p}+c^{-}_{q}\big)\delta\big(c^{+}_{p}+c^{+}_{q}\big)}{\sqrt{4pq}}\bigg\{\dfrac{V^{(2)}_{eff}\big(M\big)}{2!}\hat{\alpha}_{p}\hat{\alpha}_{q}+h.c.\bigg\}=\nonumber \\
     = \lambda^{2}\int_{0}^{+\infty}\dfrac{dp}{2\pi}\int_{0}^{+\infty}\dfrac{dq}{2\pi}\dfrac{(2\pi)^{2} \delta\big(|p|+|q|\big)\delta\big(p+q\big)}{\sqrt{4pq}}\bigg\{\dfrac{V^{(2)}_{eff}\big(M\big)}{2!}\hat{\alpha}_{p}\hat{\alpha}_{q}+h.c.\bigg\} \nonumber \\
     =\lambda^{2}\int_{0}^{+\infty}\dfrac{dp}{2\pi}\int_{0}^{+\infty}\dfrac{dq}{2\pi}\dfrac{2\pi \delta(p+q)}{\sqrt{4pq}}V_{2}(0)\bigg\{\hat{\alpha}_{p}\hat{\alpha}_{q}+h.c.\bigg\}\approx\nonumber \\
    \approx\lambda^{2}\int_{0}^{+\infty}\dfrac{dp}{2\pi}\int_{0}^{+\infty}\dfrac{dq}{2\pi}\dfrac{2\pi \delta(p+q)}{\sqrt{4pq}}V_{2}(0)\bigg\{\hat{\alpha}_{0}\hat{\alpha}_{0}+h.c.\bigg\},
\end{eqnarray}
where we have used that from \eqref{mopou5t} it follows that $V_{2}(p + q) = \frac12 \, V_{eff}^{(2)}(M) \, 2 \pi \, \delta(c_p^+ + c_q^+)$ for $\Phi = M = const$. Then, using the following representation of the delta function
\begin{equation}
    \label{6666666}
    \delta(x)=\lim_{\epsilon \to 0}\dfrac{1}{2\epsilon}e^{-|x|/\epsilon},
\end{equation}
and the fact that
\begin{equation}
    \label{6666666v}
    \int_{0}^{+\infty}dp\dfrac{1}{\sqrt{p}}e^{-p/\epsilon}=\sqrt{\pi\epsilon},
\end{equation}
we get
\begin{eqnarray}
     \label{6666666666}
     \lambda^{2}\int_{0}^{+\infty}\dfrac{dp}{2\pi}\int_{0}^{+\infty}\dfrac{dq}{2\pi}\dfrac{2\pi \delta(p+q)}{\sqrt{4pq}}=\dfrac{\lambda^{2}}{4\pi}\lim_{\epsilon \to 0}\dfrac{1}{2\epsilon}\int_{0}^{+\infty}dp\int_{0}^{+\infty}dq\dfrac{1}{\sqrt{pq}}e^{-(p+q)/\epsilon}=\dfrac{\lambda^{2}}{4\pi}\lim_{\epsilon \to 0}\dfrac{\pi \epsilon}{2\epsilon}=\dfrac{\lambda^{2}}{8}. \ \
\end{eqnarray}
Finally,
\begin{eqnarray}
     \label{666C138}
     \lambda^{2}\int_{0}^{+\infty}\dfrac{dp}{2\pi}\int_{0}^{+\infty}\dfrac{dq}{2\pi}\dfrac{2\pi \delta\big(c^{-}_{p}+c^{-}_{q}\big)}{\sqrt{4pq}}\bigg\{V_{2}(p+q)\hat{\alpha}_{p}\hat{\alpha}_{q}+h.c.\bigg\}=\dfrac{\lambda^{2}}{8}V_{2}(0)\bigg\{\hat{\alpha}_{0}\hat{\alpha}_{0}+h.c.\bigg\}.
\end{eqnarray}
The structure of this term shows that it is responsible for the presence of the condensate \cite{Popov:1984mx} and is similar to the term that would appear in the effective action due to the stationary solution for $\kappa_{pq}$, which was found in the previous section. Similar analysis shows that terms containing more than two annihilation operators will disappear due to the energy and momentum conservation in the case when $\Phi = M = const$.

In all, these observations show a sign that in non--trivial $\Phi(v)$ background the loop corrections to the anomalous quantum averages $\langle \hat{a}_{p}\hat{a}_{q}\rangle$, $\langle \hat{a}_{p}\hat{a}_{q}\hat{a}_{r}\rangle$ and etc. may show some secular behaviour, which is not present for $\Phi = const$. And in this paper we essentially see a dynamical generation of the scalar condensate.

\section{Acknowledgments}

We would like to acknowledge discussions with A.Semenov, A.Alexandrov, A.Badamshina and F.Popov.

The work of ETA was supported by the grant from the Foundation for the Advancement of Theoretical Physics and Mathematics ``BASIS'' and by RFBR grant 19-02-00815. The work was supported by Russian Ministry of education and science (project 5-100).

\appendix
\section{Calculation of one loop corrections to the Keldysh Green function for scalars} \label{A}

In this Appendix we show that when $T_{0} \rightarrow -\infty$ and $$t_{1} \approx t_{2} =T \rightarrow +\infty$$
the first term in \eqref{eq:25} which is proportional to $e^{-i(u_{3}-u_{4})(k_{1}+k_{1})-ia(3,4)(1/k_{1}+1/k_{2})}$ is vanishing. We assume that the integration over $t_{3},t_{4}$ start from $T_{0}$ -- the moment after which the self interaction $\lambda \phi \bar{\psi} \psi$ is adiabatically turned on and we shift this moment to $-\infty$ where it is possible. In what follows, we omit the integration over $k_{1}$ and $k_{2}$ and functions of those variables as well as integration over $v_{3}$ and $v_{4}$: we restore them in the end.

First, we consider the term which is proportional to $\theta(t_{2}-t_{4})\theta(t_{4}-t_{3})$ in \eqref{eq:25} under the integrals over $k_{1},k_{2},v_{3},v_{4}$ and rewrite the arguments of the Heaviside functions via the light-cone coordinates:
\begin{eqnarray}
    \label{eq:27}
    \int du_{3}\int du_{4}e^{iu_{3}c^{-}_{p}+iv_{3}c^{+}_{p}-iu_{4}c^{-}_{q}-iv_{4}c^{+}_{q}-i(u_{3}-u_{4})(k_{1}+k_{1})-ia(3,4)(1/k_{1}+1/k_{2})} \theta(T-t_{4})\theta(t_{4}-t_{3})\theta(t_{4}-T_{0})=\nonumber \\
    =\int_{T_{0}-v_{4}}^{T-v_{4}}du_{4}\int_{-\infty}^{u_{4}+v_{4}-v_{3}}du_{3}e^{iu_{3}c^{-}_{p}+iv_{3}c^{+}_{p}-iu_{4}c^{-}_{q}-iv_{4}c^{+}_{q}-i(u_{3}-u_{4})(k_{1}+k_{1})-ia(3,4)(1/k_{1}+1/k_{2})} = \bigg|u_{4}\rightarrow u_{4}-v_{4}\bigg|= \nonumber \\
    =\int_{T_{0}}^{T}du_{4}\dfrac{1}{i\big(c^{-}_{p}-(k_{1}+k_{2})-i\epsilon\big)}e^{iu_{4}(c^{-}_{p}-c^{-}_{q})+iv_{3}(c^{+}_{p}-c^{-}_{p}+k_{1}+k_{2})-iv_{4}(c^{+}_{q}-c^{-}_{q}+k_{1}+k_{2})-ia(3,4)(1/k_{1}+1/k_{2})}= \nonumber \\
    =\dfrac{2\pi}{i\big(c^{-}_{p}-(k_{1}+k_{2})-i\epsilon\big)}\delta(c^{-}_{p}-c^{-}_{q})e^{iv_{3}(c^{+}_{p}-c^{-}_{p}+k_{1}+k_{2})-iv_{4}(c^{+}_{q}-c^{-}_{q}+k_{1}+k_{2})-ia(3,4)(1/k_{1}+1/k_{2})}. \ \ \ \  \ \ \
\end{eqnarray}
Second, we consider the term which is proportional to $\theta(t_{1}-t_{3})\theta(t_{3}-t_{4})$ and rewrite the arguments of the Heaviside functions via the light-cone coordinates:
\begin{eqnarray}
  \label{eq:28}
    \int du_{3}\int du_{4}e^{iu_{3}c^{-}_{p}+iv_{3}c^{+}_{p}-iu_{4}c^{-}_{q}-iv_{4}c^{+}_{q}-i(u_{3}-u_{4})(k_{1}+k_{1})-ia(3,4)(1/k_{1}+1/k_{2})} \theta(T-t_{3})\theta(t_{3}-t_{4})\theta(t_{3}-T_{0})=\nonumber \\
    =\int_{T_{0}-v_{3}}^{T-v_{3}}du_{3}\int_{-\infty}^{u_{3}+v_{3}-v_{4}}du_{4}e^{iu_{3}c^{-}_{p}+iv_{3}c^{+}_{p}-iu_{4}c^{-}_{q}-iv_{4}c^{+}_{q}-i(u_{3}-u_{4})(k_{1}+k_{1})-ia(3,4)(1/k_{1}+1/k_{2})} = \bigg|u_{3}\rightarrow u_{3}-v_{3}\bigg|= \nonumber \\
   =\int_{T_{0}}^{T}du_{3}\dfrac{-1}{i\big(c^{-}_{q}-(k_{1}+k_{2})+i\epsilon\big)}e^{iu_{3}(c^{-}_{p}-c^{-}_{q})+iv_{3}(c^{+}_{p}-c^{-}_{p}+k_{1}+k_{2})-iv_{4}(c^{+}_{q}-c^{-}_{q}+k_{1}+k_{2})-ia(3,4)(1/k_{1}+1/k_{2})}= \nonumber \\
    =\dfrac{-2\pi}{i\big(c^{-}_{p}-(k_{1}+k_{2})+i\epsilon\big)}\delta(c^{-}_{p}-c^{-}_{q})e^{iv_{3}(c^{+}_{p}-c^{-}_{p}+k_{1}+k_{2})-iv_{4}(c^{+}_{q}-c^{-}_{q}+k_{1}+k_{2})-ia(3,4)(1/k_{1}+1/k_{2})}.
\end{eqnarray}
Third, we consider the term which is proportional to $\theta(t_{1}-t_{3})\theta(t_{2}-t_{4})$ and rewrite the arguments of the Heaviside functions via the light-cone coordinates:
\begin{eqnarray}
    \label{eq:29}
    \int du_{3}\int du_{4}e^{iu_{3}c^{-}_{p}+iv_{3}c^{+}_{p}-iu_{4}c^{-}_{q}-iv_{4}c^{+}_{q}-i(u_{3}-u_{4})(k_{1}+k_{1})-ia(3,4)(1/k_{1}+1/k_{2})} \theta(T-t_{3})\theta(T-t_{4})\theta(t_{3}-T_{0})\theta(t_{4}-T_{0})\nonumber \\
    =\int_{T_{0}-v_{3}}^{T-v_{3}}du_{3}\int_{T_{0}-v_{4}}^{T-v_{4}}du_{4}e^{iu_{3}c^{-}_{p}+iv_{3}c^{+}_{p}-iu_{4}c^{-}_{q}-iv_{4}c^{+}_{q}-i(u_{3}-u_{4})(k_{1}+k_{1})-ia(3,4)(1/k_{1}+1/k_{2})} = \bigg|u_{3(4)}\rightarrow u_{3(4)}-v_{3(4)}\bigg| \nonumber \\
    =\int_{T_{0}}^{T}du_{3}\int_{T_{0}}^{T}du_{4}e^{iu_{3}(c^{-}_{p}-(k_{1}+k_{2}))-iu_{4}(c^{-}_{q}-(k_{1}+k_{2}))+iv_{3}(c^{+}_{p}-c^{-}_{p}+k_{1}+k_{2})-iv_{4}(c^{+}_{q}-c^{-}_{q}+k_{1}+k_{2})-ia(3,4)(1/k_{1}+1/k_{2})}= \nonumber \\
    =(2\pi)^{2}\delta \big(c^{-}_{p}-(k_{1}+k_{2}) \big)\delta \big(c^{-}_{p}-c^{-}_{q}\big)e^{iv_{3}(c^{+}_{p}-c^{-}_{p}+k_{1}+k_{2})-iv_{4}(c^{+}_{q}-c^{-}_{q}+k_{1}+k_{2})-ia(3,4)(1/k_{1}+1/k_{2})}. \nonumber \\
\end{eqnarray}
It is worth mentioning here that we obtain $\delta(c^{-}_{p}-c^{-}_{q})$ only in the limit $t_{1}, t_{2}\rightarrow \infty$ and $T_{0} \rightarrow -\infty$.

Now, analyzing \eqref{eq:25}, \eqref{eq:27}, \eqref{eq:28} and \eqref{eq:29} we see that the first term in \eqref{eq:25} vanishes:
\begin{eqnarray}
    \label{7}
    \bigg(-\theta(t_{2}-t_{4})\theta(t_{4}-t_{3})-\theta(t_{1}-t_{3})\theta(t_{3}-t_{4})+\theta(t_{1}-t_{3})\theta(t_{2}-t_{4})\bigg) \rightarrow \nonumber \\
    \rightarrow \bigg( -\dfrac{1}{c^{-}_{p}-(k_{1}+k_{2})-i\epsilon}+\dfrac{1}{c^{-}_{p}-(k_{1}+k_{2})+i\epsilon}+2\pi \delta \big(c^{-}_{p}-(k_{1}+k_{2}) \bigg) \nonumber \\
    \propto -2\pi \delta \big(c^{-}_{p}-(k_{1}+k_{2})\big)+2\pi \delta \big(c^{-}_{p}-(k_{1}+k_{2})=0,
\end{eqnarray}
while the second term in \eqref{eq:25} will add up and give:
\begin{eqnarray}
    \label{8}
    \bigg(\theta(t_{2}-t_{4})\theta(t_{4}-t_{3})+\theta(t_{1}-t_{3})\theta(t_{3}-t_{4})+\theta(t_{1}-t_{3})\theta(t_{2}-t_{4})\bigg) \rightarrow  \nonumber \\
    \rightarrow 2\pi \delta \big(c^{-}_{p}+(k_{1}+k_{2})\big)+2\pi \delta \big(c^{-}_{p}+(k_{1}+k_{2})\big)
\end{eqnarray}
and,finally, we get:
\begin{eqnarray}
    \label{eq:30}
    2n^{(2)}_{qp} \approx\dfrac{i(2\pi)^{2}}{\sqrt{2|p|2|q|}}\int_{0}^{+\infty}\dfrac{dk_{1}}{2\pi}\int_{0}^{+\infty}\dfrac{dk_{2}}{2\pi}\bigg(\dfrac{k_{1}-k_{2}}{k_{1}k_{2}}\bigg)^{2}\delta \big(c^{-}_{p}-c^{-}_{q}\big)\delta \big(c^{-}_{p}+(k_{1}+k_{2})\big) \times \nonumber \\
    \times \int dv_{3}\int dv_{4}\Phi(v_{3})\Phi(v_{4})e^{iv_{3}c^{+}_{p}-iv_{4}c^{+}_{q}+ia(3,4)(1/k_{1}+1/k_{2})}.
\end{eqnarray}
Since in the $\kappa_{pq}$ there is only the difference between Heaviside functions, using \eqref{eq:27} and \eqref{eq:28} we will get principle values and not the delta functions:

\begin{eqnarray}
    \label{6}
 \theta(t_{2}-t_{4})\theta(t_{4}-t_{3})-\theta(t_{1}-t_{3})\theta(t_{3}-t_{4}) +... \rightarrow \nonumber \\ \rightarrow \dfrac{1}{c^{-}_{p}-(k_{1}+k_{2})-i\epsilon}+\dfrac{1}{c^{-}_{p}-(k_{1}+k_{2})+i\epsilon}=2\dfrac{c^{-}_{p}-(k_{1}+k_{2})}{\big(c^{-}_{p}-(k_{1}+k_{2})\big)^{2}+\epsilon^{2}}+...
\end{eqnarray}
Then, we get that:
\begin{eqnarray}
    \label{11}
    2\kappa^{(2)}_{pq}=\dfrac{-1}{\sqrt{2|p|2|q|}}\int_{0}^{+\infty}\dfrac{dk_{1}}{2\pi}\int_{0}^{+\infty}\dfrac{dk_{2}}{2\pi}\dfrac{1}{2}\bigg(\dfrac{k_{1}-k_{2}}{k_{1}k_{2}}\bigg)^{2}\int dv_{3}\int dv_{4}\Phi(v_{3})\Phi(v_{4})2\pi \delta \big(c^{-}_{p}+c^{-}_{q}\big) \nonumber \\
     \bigg[ e^{iv_{3}\big(c^{+}_{p}-c^{-}_{p}+k_{1}+k_{2}\big)+iv_{4}\big(c^{+}_{q}-c^{-}_{q}-(k_{1}+k_{2})\big)-ia(3,4)(1/k_{1}+1/k_{2})}\bigg(-2 \ \mathcal{P}\dfrac{1}{c^{-}_{p}-(k_{1}+k_{2})}+2\pi\delta\big(c^{-}_{p}-(k_{1}+k_{2})\big)\bigg)+ \nonumber \\
    +e^{iv_{3}\big(c^{+}_{p}-c^{-}_{p}-(k_{1}+k_{2})\big)+iv_{4}\big(c^{+}_{q}-c^{-}_{q}+(k_{1}+k_{2})\big)+ia(3,4)(1/k_{1}+1/k_{2})}\bigg(2 \ \mathcal{P}\dfrac{1}{c^{-}_{p}+(k_{1}+k_{2})}+2\pi\delta\big(c^{-}_{p}+(k_{1}+k_{2})\big)\bigg)\bigg]. \nonumber \\
\end{eqnarray}
Next, let us analyze the structure of $n^{(2)}_{qp}$ and $\kappa^{(2)}_{pq}$ for different values of momenta $p,q$.
First, when $p>0$ and $q>0$ we have that $$c^{-}_{p} \propto |p|-p=0. $$
Then
\begin{eqnarray}
    \label{eq:31}
    2n^{(2)}_{qp}=\big(T-T_{0}\big)\dfrac{i(2\pi)^{2}}{\sqrt{2|p|2|q|}}\int_{0}^{+\infty}\dfrac{dk_{1}}{2\pi}\int_{0}^{+\infty}\dfrac{dk_{2}}{2\pi}\bigg(\dfrac{k_{1}-k_{2}}{k_{1}k_{2}}\bigg)^{2}\delta \big(k_{1}+k_{2}\big) \times \nonumber \\
    \times \int dv_{3}\int dv_{4}\Phi(v_{3})\Phi(v_{4})e^{iv_{3}c^{+}_{p}-iv_{4}c^{+}_{q}+ia(3,4)(1/k_{1}+1/k_{2})} \sim 0,
\end{eqnarray}
since it takes non zero value only when $k_{1}=k_{2}=0$. But our integrals over $k_{1}$ and $k_{2}$ contain parts $\lim_{\epsilon \to 0 }e^{-\epsilon/k_{1}}$ (see the footnote 2) which will make the integrands zero even when $k_{1}=k_{2}=0$.

Similarly,
\begin{eqnarray}
    \label{12}
    2\kappa^{(2)}_{pq} \approx \bigg(T-T_{0}\bigg)\dfrac{-1}{\sqrt{2|p|2|q|}}\int_{0}^{+\infty}\dfrac{dk_{1}}{2\pi}\int_{0}^{+\infty}\dfrac{dk_{2}}{2\pi}\dfrac{1}{2}\bigg(\dfrac{k_{1}-k_{2}}{k_{1}k_{2}}\bigg)^{2}\int dv_{3}\int dv_{4}\Phi(v_{3})\Phi(v_{4}) \times \nonumber \\
    \times \bigg[ e^{iv_{3}\big(\sqrt{2}p+k_{1}+k_{2}\big)+iv_{4}\big(\sqrt{2}q-(k_{1}+k_{2})\big)-ia(3,4)(1/k_{1}+1/k_{2})}\bigg(2 \ \mathcal{P}\dfrac{1}{(k_{1}+k_{2})}\bigg)+ \nonumber \\
    +e^{iv_{3}\big(\sqrt{2}p-(k_{1}+k_{2})\big)+iv_{4}\big(\sqrt{2}q+(k_{1}+k_{2})\big)+ia(3,4)(1/k_{1}+1/k_{2})}\bigg(2 \ \mathcal{P}\dfrac{1}{(k_{1}+k_{2})}\bigg)\bigg],
\end{eqnarray}
where we used that
\begin{equation*}
    2\pi\delta \big(c^{-}_{p}+c^{-}_{q}\big) \approx \int_{T_{0}}^{T}du e^{iu\big(c^{-}_{p}+c^{-}_{q}\big)}, \qquad \text{as} \qquad T-T_{0}\rightarrow \infty
\end{equation*}
from which it follows that $$2\pi\delta(0)=T-T_{0}.$$
Second, when $p>0$ and $q<0$ we have
\begin{eqnarray}
     \label{eq:32}
    2n^{(2)}_{qp} \approx \delta \big(\sqrt{2}q\big)\dfrac{i(2\pi)^{2}}{\sqrt{2|p|2|q|}}\int_{0}^{+\infty}\dfrac{dk_{1}}{2\pi}\int_{0}^{+\infty}\dfrac{dk_{2}}{2\pi}\bigg(\dfrac{k_{1}-k_{2}}{k_{1}k_{2}}\bigg)^{2}\delta \big(k_{1}+k_{2}\big) \times \nonumber \\
    \times \int dv_{3}\int dv_{4}\Phi(v_{3})\Phi(v_{4})e^{iv_{3}c^{+}_{p}-iv_{4}c^{+}_{q}+ia(3,4)(1/k_{1}+1/k_{2})} \approx 0,
\end{eqnarray}
and
\begin{eqnarray}
    \label{13}
    2\kappa^{(2)}_{pq}\approx 2\pi \delta \big(\sqrt{2}q\big)\dfrac{-1}{\sqrt{2|p|2|q|}}\int_{0}^{+\infty}\dfrac{dk_{1}}{2\pi}\int_{0}^{+\infty}\dfrac{dk_{2}}{2\pi}\dfrac{1}{2}\bigg(\dfrac{k_{1}-k_{2}}{k_{1}k_{2}}\bigg)^{2}\int dv_{3}\int dv_{4}\Phi(v_{3})\Phi(v_{4})\times \nonumber \\
    \times \bigg[ e^{iv_{3}\big(\sqrt{2}p+k_{1}+k_{2}\big)+iv_{4}\big(-(k_{1}+k_{2})\big)-ia(3,4)(1/k_{1}+1/k_{2})}\bigg(2 \ \mathcal{P}\dfrac{1}{(k_{1}+k_{2})}+2\pi\delta\big((k_{1}+k_{2})\big)\bigg)+ \nonumber \\
    +e^{iv_{3}\big(\sqrt{2}p-(k_{1}+k_{2})\big)+iv_{4}\big((k_{1}+k_{2})\big)+ia(3,4)(1/k_{1}+1/k_{2})}\bigg(2 \ \mathcal{P}\dfrac{1}{(k_{1}+k_{2})}+2\pi\delta\big((k_{1}+k_{2})\big)\bigg)\bigg],
\end{eqnarray}
which is not equal to zero only when $q=0$.

Third, when $p<0$ and $q>0$ we have
\begin{eqnarray}
     \label{eq:33}
    2n^{(2)}_{qp} \approx \delta \big(\sqrt{2}p\big)\dfrac{i(2\pi)^{2}}{\sqrt{2|p|2|q|}}\int_{0}^{+\infty}\dfrac{dk_{1}}{2\pi}\int_{0}^{+\infty}\dfrac{dk_{2}}{2\pi}\bigg(\dfrac{k_{1}-k_{2}}{k_{1}k_{2}}\bigg)^{2}\delta \big(k_{1}+k_{2}\big) \times \nonumber \\
    \times \int_{}^{} dv_{3}\int_{}^{} dv_{4}\Phi(v_{3})\Phi(v_{4})e^{iv_{3}c^{+}_{p}-iv_{4}c^{+}_{q}+ia(3,4)(1/k_{1}+1/k_{2})}\approx 0,
\end{eqnarray}
and
\begin{eqnarray}
    \label{14}
    2\kappa^{(2)}_{pq} \approx 2\pi \delta \big(\sqrt{2}p\big)\dfrac{-1}{\sqrt{2|p|2|q|}}\int_{0}^{+\infty}\dfrac{dk_{1}}{2\pi}\int_{0}^{+\infty}\dfrac{dk_{2}}{2\pi}\dfrac{1}{2}\bigg(\dfrac{k_{1}-k_{2}}{k_{1}k_{2}}\bigg)^{2}\int dv_{3}\int dv_{4}\Phi(v_{3})\Phi(v_{4})\times \nonumber \\
    \times \bigg[ e^{iv_{3}\big(k_{1}+k_{2}\big)+iv_{4}\big(\sqrt{2}q-(k_{1}+k_{2})\big)-ia(3,4)(1/k_{1}+1/k_{2})}\bigg(2 \ \mathcal{P}\dfrac{1}{(k_{1}+k_{2})}+2\pi\delta\big((k_{1}+k_{2})\big)\bigg)+ \nonumber \\
    +e^{iv_{3}\big(-(k_{1}+k_{2})\big)+iv_{4}\big(\sqrt{2}q+(k_{1}+k_{2})\big)+ia(3,4)(1/k_{1}+1/k_{2})}\bigg(2 \ \mathcal{P}\dfrac{1}{k_{1}+k_{2}}+2\pi\delta \big((k_{1}+k_{2})\big)\bigg)\bigg].
\end{eqnarray}
which is not equal to zero only when $p=0$.

Finally, when $p<0$ and $q<0$ we obtain that $$c^{+}_{p} \propto |p|+p=0. $$ Then
\begin{eqnarray}
    \label{eq:34}
    2n^{(2)}_{qp} \approx \dfrac{1}{\sqrt{2}}\delta \big(p-q\big)\dfrac{i(2\pi)^{2}}{\sqrt{2|p|2|q|}}\int_{0}^{+\infty}\dfrac{dk_{1}}{2\pi}\int_{0}^{+\infty}\dfrac{dk_{2}}{2\pi}\bigg(\dfrac{k_{1}-k_{2}}{k_{1}k_{2}}\bigg)^{2}\delta \big(-\sqrt{2}p+(k_{1}+k_{2})\big) \times \nonumber \\
    \times \int dv_{3}\int dv_{4}\Phi(v_{3})\Phi(v_{4})e^{ia(3,4)(1/k_{1}+1/k_{2})}\approx 0,
\end{eqnarray}
and
\begin{eqnarray}
    \label{15}
    2\kappa^{(2)}_{pq}\approx 2\pi \delta \big(\sqrt{2}p+\sqrt{2}q\big)\dfrac{-1}{\sqrt{2|p|2|q|}}\int_{0}^{+\infty}\dfrac{dk_{1}}{2\pi}\int_{0}^{+\infty}\dfrac{dk_{2}}{2\pi}\dfrac{1}{2}\bigg(\dfrac{k_{1}-k_{2}}{k_{1}k_{2}}\bigg)^{2}\int dv_{3}\int dv_{4}\Phi(v_{3})\Phi(v_{4})\times \nonumber \\
    \times \bigg[ e^{iv_{3}\big(\sqrt{2}p+k_{1}+k_{2}\big)+iv_{4}\big(\sqrt{2}q-(k_{1}+k_{2})\big)-ia(3,4)(1/k_{1}+1/k_{2})}\bigg(2 \ \mathcal{P}\dfrac{1}{\sqrt{2}p+(k_{1}+k_{2})}\bigg)+ \nonumber \\
    +e^{iv_{3}\big(\sqrt{2}p-(k_{1}+k_{2})\big)+iv_{4}\big(\sqrt{2}q+(k_{1}+k_{2})\big)+ia(3,4)(1/k_{1}+1/k_{2})}\bigg(2 \ \mathcal{P}\dfrac{1}{-\sqrt{2}p+(k_{1}+k_{2})}\bigg)+\nonumber \\
    + e^{iv_{3}\big(\sqrt{2}p+k_{1}+k_{2}\big)+iv_{4}\big(\sqrt{2}q-(k_{1}+k_{2})\big)-ia(3,4)(1/k_{1}+1/k_{2})}\bigg(2\pi\delta\big(-\sqrt{2}p-(k_{1}+k_{2})\big)\bigg)+ \nonumber \\
    +e^{iv_{3}\big(\sqrt{2}p-(k_{1}+k_{2})\big)+iv_{4}\big(\sqrt{2}q+(k_{1}+k_{2})\big)+ia(3,4)(1/k_{1}+1/k_{2})}\bigg(2\pi\delta\big(-\sqrt{2}p+(k_{1}+k_{2})\big)\bigg)\bigg]. \ \  \ \ \
\end{eqnarray}
which is not zero only when $p=q=0$.

In all, we get that
\begin{equation}
    \label{16}
    n^{(2)}_{qp}=0, \qquad  \text{for any values of $p,q$.}
\end{equation}
And $\kappa_{pq}$ has non zero value only when $p \geq 0$ and $q \geq 0$ which is equal to
\begin{eqnarray}
    \label{17}
    2\kappa^{(2)}_{pq}=\bigg(T-T_{0}\bigg)\dfrac{-1}{\sqrt{2|p|2|q|}}\int_{0}^{+\infty}\dfrac{dk_{1}}{2\pi}\int_{0}^{+\infty}\dfrac{dk_{2}}{2\pi}\dfrac{1}{2}\bigg(\dfrac{k_{1}-k_{2}}{k_{1}k_{2}}\bigg)^{2}\int dv_{3}\int dv_{4}\Phi(v_{3})\Phi(v_{4}) \times \nonumber \\
    \times \bigg[ e^{iv_{3}\big(\sqrt{2}p+k_{1}+k_{2}\big)+iv_{4}\big(\sqrt{2}q-(k_{1}+k_{2})\big)-ia(3,4)(1/k_{1}+1/k_{2})}\bigg(2 \ \mathcal{P}\dfrac{1}{(k_{1}+k_{2})}\bigg)+ \nonumber \\
    +e^{iv_{3}\big(\sqrt{2}p-(k_{1}+k_{2})\big)+iv_{4}\big(\sqrt{2}q+(k_{1}+k_{2})\big)+ia(3,4)(1/k_{1}+1/k_{2})}\bigg(2 \ \mathcal{P}\dfrac{1}{(k_{1}+k_{2})}\bigg)\bigg].
\end{eqnarray}

\section{Derivation of the Dyson-Schwinger equation for the scalar Keldysh propagator} \label{B}
In the Appendix we derive the Dyson-Schwinger equation for $k_{pq}$. We start from the eq.\eqref{eQ:107}. Using the tree-level propagators for fermions \eqref{eq:18}-\eqref{eq:20} we get
\begin{eqnarray}
    \label{eq:108}
    \kappa_{pq}(t_{1},t_{2})=\kappa^{(2)}_{pq}+\nonumber \\
    +\dfrac{1}{2}\bigg\{ i\int_{-\infty}^{+\infty}\dfrac{dq_{1}}{2\pi}\dfrac{1}{\sqrt{2|q_{1}|}}\dfrac{1}{\sqrt{2|p|}}\int_{0}^{+\infty}\dfrac{dk_{1}}{2\pi}\int_{0}^{+\infty}\dfrac{dk_{2}}{2\pi}\dfrac{1}{2}\bigg(\dfrac{k_{1}-k_{2}}{k_{1}k_{2}}\bigg)^{2}\int dv_{3}\int dv_{4}\Phi(v_{3})\Phi(v_{4})\kappa_{q_{1}q}(t_{4},t_{2})\bigg[\nonumber \\
    \int_{T_{0}}^{t_{1}}du_{3} \int_{-\infty}^{u_{3}-v_{4}}du_{4}e^{i(u_{3}-v_{3})c^{-}_{p}+iv_{3}c^{+}_{p}-iu_{4}c^{-}_{q_{1}}-iv_{4}c^{+}_{q_{1}}}\bigg(e^{i(k_{1}+k_{2})(u_{3}-v_{3}-u_{4})+ia(v_{3},v_{4})(1/k_{1}+1/k_{2})}-c.c\bigg)\bigg]+(p \leftrightarrow q)\bigg\}. \nonumber \\
\end{eqnarray}
Next, we assume that $\kappa_{pq}(t_{4},t_{2})\approx \kappa_{pq}\big((t_{4}+t_{2})/2\big)$ and is very slow function of time in comparison with oscillation modes, i.e.  $$\kappa_{q_{1}q}(t_{4},t_{2}) \approx \kappa_{q_{1}q}(T).$$
This is so called kinetic approximation \cite{20}.

After such a simplification it is possible to take integrals over $u_{3}$ and $u_{4}$ to get:
\begin{eqnarray}
    \label{eq:109}
    \int_{T_{0}}^{t_{1}}du_{3}\int_{-\infty}^{u_{3}-v_{4}}du_{4}e^{i(u_{3}-v_{3})c^{-}_{p}+iv_{3}c^{+}_{p}-iu_{4}c^{-}_{q_{1}}-iv_{4}c^{+}_{q_{1}}}\bigg(e^{i(k_{1}+k_{2})(u_{3}-v_{3}-u_{4})+ia(v_{3},v_{4})(1/k_{1}+1/k_{2})}-c.c\bigg)\kappa_{q_{1}q}(T)= \nonumber \\
    =2\pi i \delta\big(c^{-}_{p}-c^{-}_{q_{1}}\big) \bigg[\dfrac{1}{c^{-}_{q_{1}}+k_{1}+k_{2}+i\epsilon}e^{iv_{3}(c^{+}_{p}-c^{-}_{p}-k_{1}-k_{2})-iv_{4}(c^{+}_{q_{1}}-c^{-}_{q_{1}}-k_{1}-k_{2})+ia(v_{3},v_{4})(1/k_{1}+1/k_{2})}-\nonumber \\
    -\dfrac{1}{c^{-}_{q_{1}}-(k_{1}+k_{2})+i\epsilon}e^{iv_{3}(c^{+}_{p}-c^{-}_{p}+k_{1}+k_{2})-iv_{4}(c^{+}_{q_{1}}-c^{-}_{q_{1}}+k_{1}+k_{2})-ia(v_{3},v_{4})(1/k_{1}+1/k_{2})}\bigg]\kappa_{q_{1}q}(T). \ \ \
\end{eqnarray}
First, let us consider the case when  $p \geq 0$.
In such a case $\delta\big(c^{-}_{p}-c^{-}_{q_{1}}\big)=\delta\big(c^{-}_{q_{1}}\big)$ and it has non zero value\footnote{For positive values of $p$ and $q_{1}$ we have that $2\pi \delta\big(c^{-}_{p}-c^{-}_{q_{1}}\big)=2\pi\delta(0)=T-T_{0}$.} only for positive momentum $q_{1}$ and eq.  \eqref{eq:108} acquires the form
\begin{eqnarray}
    \label{eq:112}
    \kappa_{pq}(T)=\kappa^{(2)}_{pq}-\nonumber \\-
      \bigg\{\dfrac{\big(T-T_{0}\big)}{2}\int_{0}^{+\infty}\dfrac{dq_{1}}{2\pi}\dfrac{1}{\sqrt{2|q_{1}|}}\dfrac{1}{\sqrt{2|p|}}\int_{0}^{+\infty}\dfrac{dk_{1}}{2\pi}\int_{0}^{+\infty}\dfrac{dk_{2}}{2\pi}\dfrac{1}{2}\bigg(\dfrac{k_{1}-k_{2}}{k_{1}k_{2}}\bigg)^{2}\int_{}^{} dv_{3}\int_{}^{} dv_{4}\Phi(v_{3})\Phi(v_{4})\bigg[
      \nonumber \\
    \mathcal{P}\bigg(\dfrac{1}{k_{1}+k_{2}}\bigg)\bigg[e^{iv_{3}(\sqrt{2}p-k_{1}-k_{2})-iv_{4}(\sqrt{2}q_{1}-k_{1}-k_{2})+i a(v_{3},v_{4})(1/k_{1}+1/k_{2})}+ \nonumber \\
    +e^{iv_{3}(\sqrt{2}p+k_{1}+k_{2})-iv_{4}(\sqrt{2}q_{1}+k_{1}+k_{2})-i a(v_{3},v_{4})(1/k_{1}+1/k_{2})}\bigg]\kappa_{q_{1}q}(T) + (p \leftrightarrow q)\bigg\} \ \ \
\end{eqnarray}
Now, using the exact form of $\kappa^{(2)}_{pq}$ in \eqref{tpp3} we get the following equation
\begin{equation}
    \label{eq:117}
    \kappa_{pq}(T)=(T-T_{0})F(p,q)\theta(q)+\dfrac{1}{2}\bigg\{(T-T_{0})\int_{0}^{+\infty}\dfrac{dq_{1}}{2\pi}F(p,-q_{1})\kappa_{q_{1}q}(T) + (p \leftrightarrow q)\bigg\},
\end{equation}
where $F(p,q)$ is defined in \eqref{eq:tpp4}.
Since in our approximation $\kappa_{pq}$ is a very slow function of $T$ we can also rewrite equation \eqref{eq:117} as

\begin{equation}
    \label{eq:119}
    \dfrac{\kappa_{pq}-0}{T-T_{0}} \approx \partial_{T}\kappa_{pq}(T)=F(p,q)\theta(q)+\dfrac{1}{2}\bigg\{\int_{0}^{+\infty}\dfrac{dq_{1}}{2\pi}F(p,-q_{1})\kappa_{q_{1}q}(T)+(p \leftrightarrow q)\bigg\},
\end{equation}
where we used that $\kappa^{0}_{pq}=0$.

In the case when $p<0$ we get that $\delta\big(c^{-}_{p}-c^{-}_{q_{1}}\big)=\delta\big(\sqrt{2}(p-q_{1})\big)$ for negative values of $q_{1}$ and $\delta\big(c^{-}_{p}-c^{-}_{q_{1}}\big)=\delta\big(\sqrt{2}p\big)$ for positive values of momentum $q_{1}$. Finally, using the fact that $\kappa^{(2)}_{pq}$ is zero for negative vlues of $p$ we get that equation  \eqref{eq:108} becomes
\begin{eqnarray}
     \label{eq:125}
    \kappa_{pq}(T)=\dfrac{1}{2}\int_{-\infty}^{0}\dfrac{dq_{1}}{2\pi} \delta(p-q_{1})L(p)\kappa_{q_{1}q}(T)+(p\leftrightarrow q)=\dfrac{L(p)+L(q)}{2}\kappa_{pq}(T),
\end{eqnarray}
where $L(p)$ can be found from \eqref{eq:108}.
This means that
\begin{equation}
    \label{eqQ:127}
    \kappa_{pq}(T)=0, \ \ \text{if} \ \ p<0.
\end{equation}
Finally, if we use the symmetry  over $p$, $q$ momentum, the Dyson-Schwinger equation for the anomalous average $\kappa_{pq}$ has the form

\begin{equation}
    \label{eq:126}
    \partial_{T}\kappa_{pq}(T)=F(p,q)+\dfrac{1}{2}\bigg\{\int_{0}^{+\infty}\dfrac{dq_{1}}{2\pi}F(p,-q_{1})\kappa_{q_{1}q}(T) +(p \leftrightarrow q)\bigg\}, \qquad p,q\geq0 
\end{equation}

\section{Some comments about one loop corrections to the Keldysh propagator for scalars in the case of constant classical background field}\label{C}

As was discussed in \cite{12} the exact fermion modes \eqref{eq:15} in case when $\Phi(v)=m$ define the same Fock ground state as the usual plane wave modes for massive fermions. It means that if we perform the following change of variables in \eqref{eq:30} and \eqref{11}:

\begin{equation*}
    \begin{cases}
    k_{1} \rightarrow \dfrac{1}{\sqrt{2}}(w_{k_{1}}-k_{1}) \\
    k_{2} \rightarrow \dfrac{1}{\sqrt{2}}(w_{k_{2}}-k_{2})
    \end{cases}, \qquad \qquad \text{where} \qquad w_{k}=\sqrt{m^{2}+k^{2}}.
\end{equation*}
our expressions for $n^{(2)}_{qp}$ and $\kappa^{(2)}_{pq}$ must coincide with those for the usual Fock space ground state for the plane wave modes. Indeed, this is the case. Such a calculation we perform as an independent check that our calculations do not contain mistakes.

\section{Calculation of loop corrections in 2+1 dimensions}\label{3D}
In 2+1 dimensions we can work with two dimensional spinors, which are solutions of the Dirac equation
\begin{equation}
    \label{eq:Y1}
    \bigg(i\gamma^{\mu}\partial_{\mu}-\Phi(v)\bigg)\psi=0,
    \end{equation}
with gamma matrices as
\begin{equation}
    \label{eq:Y2}
    \gamma^{0}=\left[ \begin{array}{cc}
   0 & -i \\
    i & 0 \\
  \end{array} \right],  \qquad  \gamma^{1}=\left[ \begin{array}{cc}
   0 & i \\
    i & 0 \\
  \end{array} \right],  \qquad  \gamma^{2}=\left[ \begin{array}{cc}
   i & 0 \\
    0 & -i \\
  \end{array} \right].
\end{equation}
Rewriting \eqref{eq:Y1} more explicitly, we obtain
\begin{equation}
    \label{eq:Y3}
  \left[ {\begin{array}{cc}
   -\Phi(v)-\partial_{2} & \sqrt{2}\partial_{v} \\
    \sqrt{2}\partial_{u} & -\Phi(v)+\partial_{2} \\
  \end{array} } \right]\begin{bmatrix}\psi_{1} \\ \psi_{2} \end{bmatrix}=0 \ , \qquad \begin{cases}
  v=\dfrac{t-x}{\sqrt{2}} \\
  u=\dfrac{t+x}{\sqrt{2}}
  \end{cases}.
\end{equation}
The field operator has the form
\begin{equation}
    \label{eq:Y7}
    \hat{\psi}(u,v,y)=\int_{0}^{+\infty}\dfrac{dq}{2\pi}\int_{-\infty}^{+\infty}\dfrac{dq_{2}}{2\pi}\dfrac{1}{\sqrt[4]{2}}\bigg[\hat{a}_{\vec{q}}\begin{bmatrix}\dfrac{\Phi(v)-iq_{2}}{\sqrt{2}iq} \\ 1 \end{bmatrix}e^{-iqu-i\tilde{a}(v,0)/q+iq_{2}y}+\hat{b}_{\vec{q}}^{\dagger}\begin{bmatrix}\dfrac{\Phi(v)+iq_{2}}{\sqrt{2}iq} \\ -1 \end{bmatrix}e^{iqu+i\tilde{a}(v,0)/q-iq_{2}y}\bigg] \ ,
\end{equation}
where $\Vec{q}=(q,q_{2})$ and
\begin{equation}
    \label{eq:Y5}
    \tilde{a}(v,0)=\dfrac{1}{2}\int_{0}^{v}ds\bigg(\Phi^{2}(s)+q^{2}_{2}\bigg).
\end{equation}
If one puts $\Phi(v)=m$ the standard theory of massive fermions, which does not mix positive and negative modes, follows after a trivial Bogolyubov transformation.

For scalars we use the standard plane wave expansion,
where the modes are \\ $f_{\vec{p}}(t,\vec{x}) = \dfrac{1}{\sqrt{2|p|}}e^{-i|p|t+ip_{1}x+ip_{2}y}$ with $|p|=\sqrt{p^{2}_{1}+p^{2}_{2}}$ and the ladder operators $\hat{\alpha}_{\vec{p}}$, $\hat{\alpha}_{\vec{p}}^{\dagger}$  satisfy the standard commutation relations.

The calculation of loop corrections to $\kappa^{(2)}_{\vec{p}\vec{q}}$ and $n^{(2)}_{\vec{p}\vec{q}}$ to the Keldysh propagator for scalars are almost identical to the two--dimensional case above. Hence, we present here only the result in the limit $T=\dfrac{t_{1}+t_{2}}{2} \gg |t_{1}-t_{2}|$:
\begin{equation}
    \label{Y44}
    \begin{cases}
    n^{(2)}_{\vec{q}\vec{p}} \sim 0 \\
    \kappa^{(2)}_{\vec{p}\vec{q}} \approx (T-T_{0})\delta(p_{2}+q_{2})\tilde{F}(p,q)\theta(p_{1})\theta(q_{1}), \ p_{1},q_{1},p_{2},q_{2}\geq 0.
    \end{cases},
\end{equation}
where
\begin{eqnarray}
    \label{Y42}
    \tilde{F}(p_{1},q_{1})=\dfrac{-1}{\sqrt{2|p|2|q|}}\int_{0}^{+\infty}\dfrac{dk}{2\pi}\int_{0}^{+\infty}\dfrac{dk^{\prime}}{2\pi}\int_{-\infty}^{+\infty}dk_{2} \int dv_{3}\int dv_{4} 2 \ \mathcal{P}\dfrac{1}{k+k^{\prime}}\nonumber \\
     \bigg[ A(v_{3},v_{4};k,k_{2};k^{\prime},k^{\prime}_{2}=-k_{2}+p_{2})e^{iv_{3}\big(c^{+}_{p}-c^{-}_{p}+k+k^{\prime}\big)+iv_{4}\big(c^{+}_{q}-c^{-}_{q}-(k+k^{\prime})\big)-i\tilde{a}(3,4)(1/k+1/k^{\prime})}+ \nonumber \\
    +A^{*}(v_{3},v_{4};k,k_{2};k^{\prime},k^{\prime}_{2}=-k_{2}-p_{2})e^{iv_{3}\big(c^{+}_{p}-c^{-}_{p}-(k+k^{\prime})\big)+iv_{4}\big(c^{+}_{q}-c^{-}_{q}+(k+k^{\prime})\big)+i\tilde{a}(3,4)(1/k+1/k^{\prime})}\bigg],
\end{eqnarray}
and
\begin{eqnarray}
    \label{eq:Y33}
    A(v_{3},v_{4};k,k_{2};k^{\prime},k^{\prime}_{2})=-\dfrac{(\Phi(v_{3})-ik_{2})(\Phi(v_{4})+ik^{\prime}_{2})}{2kk^{\prime}}+\dfrac{(\Phi(v_{3})-ik_{2})(\Phi(v_{4})+ik_{2})}{2k^{2}}+\nonumber\\
    +\dfrac{(\Phi(v_{4})+ik^{\prime}_{2})(\Phi(v_{3})-ik^{\prime}_{2})}{2k^{\prime 2}}-\dfrac{(\Phi(v_{4})+ik_{2})(\Phi(v_{3})-ik^{\prime}_{2})}{2kk^{\prime}}.
\end{eqnarray}
with $c^{\pm}_{p} \equiv  \dfrac{1}{\sqrt{2}}\big(|p| \pm p_{1}\big)=\dfrac{1}{\sqrt{2}}\big(\sqrt{p^{2}_{1}+p^{2}_{2}} \pm p_{1}\big)$.

Furthermore, the stationary solution of the Dyson-Schwinger equation has the form
\begin{equation}
    \label{Y46}
    \kappa_{\vec{p}\vec{q}}=-2\pi\delta(p_{2}+q_{2})\delta(p_{1}+q_{1}), \quad p_{1}, q_{1},p_{2}, q_{2} \geq 0.
\end{equation}
Now, let us calculate its contribution to the exact Keldysh function in the limit in question:

\begin{eqnarray}
    \label{eq:Y48}
    D^{K}(t_{1},\vec{x}_{1};t_{2},\vec{x}_{2}) \approx D_{0}^{K}(t_{1}-t_{2},\vec{x}_{1}-x_{2}) +\nonumber\\
    +\int_{0}^{+\infty}\dfrac{d^{2}\vec{p}}{(2\pi)^{2}}\dfrac{d^{2}\vec{q}}{(2\pi)^{2}}2\kappa_{\vec{p}\vec{q}}\dfrac{1}{\sqrt{2|p|}}e^{-i|p|t_{1}+ip_{1}x_{1}+ip_{2}y_{1}}\dfrac{1}{\sqrt{2|q|}}e^{-i|q|t_{2}+iq_{1}x_{2}+iq_{2}y_{2}} + c.c..
\end{eqnarray}
Since the static solution $\kappa_{\vec{p}\vec{q}}$ is not zero only when $\vec{p}=\vec{q}=0$ we have to carefully calculate the last expression:
\begin{eqnarray}
    \label{eq:Y47}
    \int_{0}^{+\infty}\dfrac{d^{2}\vec{p}}{(2\pi)^{2}}\dfrac{d^{2}\vec{q}}{(2\pi)^{2}}2\kappa_{\vec{p}\vec{q}}\dfrac{1}{\sqrt{2|p|2|q|}}e^{-i|p|t_{1}+ip_{1}x_{1}+ip_{2}y_{1}}e^{-i|q|t_{2}+iq_{1}x_{2}+iq_{2}y_{2}}+c.c.=\nonumber \\
    =\int_{0}^{+\infty}\dfrac{dp_{1}}{2\pi}\dfrac{dp_{2}}{2\pi}\int_{0}^{+\infty}\dfrac{dq_{1}}{2\pi}\dfrac{dq_{2}}{2\pi}\big(-2\pi \delta(p_{1}+q_{1})\delta(p_{2}+q_{2})\big)\dfrac{1}{\sqrt{|pq|}}e^{-i|p|t_{1}+ip_{1}x_{1}+ip_{2}y_{1}}e^{-i|q|t_{2}+iq_{1}x_{2}+iq_{2}y_{2}} + c.c . \nonumber\\
\end{eqnarray}
We  make the following change of variables
\begin{equation}
    \label{eq:Y49}
    \begin{cases}
     \dfrac{p_{1}+q_{1}}{\sqrt{2}}=w_{+} \\
     \dfrac{p_{1}-q_{1}}{\sqrt{2}}=w_{-}
    \end{cases}, \qquad \qquad
    \begin{cases}
     \dfrac{p_{2}+q_{2}}{\sqrt{2}}=z_{+} \\
     \dfrac{p_{2}-q_{2}}{\sqrt{2}}=z_{-}
    \end{cases}
\end{equation}
and \eqref{eq:Y47} acquires the form
\begin{eqnarray}
    \label{eq:Y50}
    -\dfrac{1}{(2\pi)^{3}}\int_{0}^{+\infty}dw_{+}\delta(w_{+})\int_{0}^{+\infty}dz_{+}\delta(z_{+})\int_{-\infty}^{+\infty}dw_{-}\int_{-\infty}^{+\infty}dz_{-}\dfrac{e^{-i\sqrt{w^{2}_{-}+z^{2}_{-}}T+iw_{-}X+iz_{-}Y}}{\sqrt{w^{2}_{-}+z^{2}_{-}}}+c.c=\nonumber\\
    =-\dfrac{1}{(2\pi)^{3}}\bigg(\dfrac{1}{2}\bigg)^{2}\int_{-\infty}^{+\infty}dw\int_{-\infty}^{+\infty}dz\dfrac{1}{\sqrt{w^{2}+z^{2}}}e^{-i\sqrt{w^{2}+z^{2}}T+iwX+izY}+c.c=\nonumber\\
    =-\dfrac{1}{(2\pi)^{3}}\bigg(\dfrac{1}{2}\bigg)^{2}2^{3}\int_{0}^{+\infty}dw\int_{0}^{+\infty}dz\dfrac{\cos{\sqrt{w^{2}+z^{2}}T}}{\sqrt{w^{2}+z^{2}}}\cos{wX}\cos{zY}, \ \
\end{eqnarray}
where
\begin{equation}
    \label{eq:Y51}
    \begin{cases}
    T=\dfrac{t_{1}+t_{2}}{\sqrt{2}}\\
    X=\dfrac{x_{1}-x_{2}}{\sqrt{2}}\\
    Y=\dfrac{y_{1}-y_{2}}{\sqrt{2}}
    \end{cases}
\end{equation}
First of all we obtain that
\begin{eqnarray}
    \label{eq:Y52}
    \int_{0}^{+\infty}dw\dfrac{\cos{\sqrt{w^{2}+z^{2}}T}}{\sqrt{w^{2}+z^{2}}}\cos{wX}=\begin{cases}
    K_{0}\bigg(z\sqrt{X^{2}-T^{2}}\bigg), \ \  \ \ |X| > |T|\\
    -\dfrac{\pi}{2}Y_{0}\bigg(z\sqrt{T^{2}-X^{2}}\bigg), \ \ |T|>|X|
    \end{cases}.
\end{eqnarray}
Then, using that \cite{bb}
\begin{eqnarray}
    \label{eq:Y53}
    \int_{0}^{+\infty}dzK_{0}\bigg(z\sqrt{X^{2}-T^{2}}\bigg)\cos{zY}=\dfrac{\pi}{2}\dfrac{1}{\sqrt{X^{2}+Y^{2}-T^{2}}}
\end{eqnarray}
and \cite{bb}
\begin{equation}
    \label{eq:Y56}
    \int_{0}^{+\infty}dzY_{0}\bigg(z\sqrt{T^{2}-X^{2}}\bigg)\cos{zY}=\begin{cases}
    -\dfrac{1}{\sqrt{X^{2}+Y^{2}-T^{2}}}, \ \sqrt{X^{2}-T^{2}}<|Y|\\
    0, \ \ \ \ \ \ \ \ \ \ \ \ \ \ \ \ \ \ \ \ \ \ \  \ \sqrt{X^{2}-T^{2}}>|Y|
    \end{cases}
\end{equation}
we find that the resummed Keldysh propagator has the following form in the limit $T\to \infty$:
\begin{eqnarray}
    \label{eq:Y59}
     D^{K}(t_{1},t_{2}; \vec{x_{1}},\vec{x_{2}}) \approx D_{0}^{K}(t_{1}-t_{2};\vec{x_{1}}-\vec{x_{2}})-\dfrac{1}{4\pi^{2}}\dfrac{\theta(|X|-|T|)+\theta(|T|-|X|)\theta\big(|Y|-\sqrt{T^{2}-X^{2}}\big)}{\sqrt{X^{2}+Y^{2}-T^{2}}}. \nonumber \\
\end{eqnarray}
This expression shows that for large distances, when $D_0^K$ vanishes (which is true in three dimensions unlike the two--dimensional case) we obtain a non--zero time $T = (t_1+t_2)/2$ dependent value, which breaks the Poincar\'e invariance.




\begin{thebibliography}{3}

\bibitem{20} L. D. Landau and E. M. Lifshitz, Vol. 10 (Pergamon Press, Oxford, 1975).

\bibitem{16} A. Kamenev, “Many-body theory of non-equilibrium systems,” Cambridge, UK: Univ. Pr. (2011)
[arXiv:cond-mat/041229].

\bibitem{2} 
  E.~T.~Akhmedov,
  Int.\ J.\ Mod.\ Phys.\ D {\bf 23}, 1430001 (2014)
  doi:10.1142/S0218271814300018
  [arXiv:1309.2557 [hep-th]].

\bibitem{1} 
  D.~Krotov and A.~M.~Polyakov,
  Nucl.\ Phys.\ B {\bf 849}, 410 (2011)
  doi:10.1016/j.nuclphysb.2011.03.025
  [arXiv:1012.2107 [hep-th]].

\bibitem{3} 
  E.~T.~Akhmedov and F.~Bascone,
  Phys.\ Rev.\ D {\bf 97}, no. 4, 045013 (2018)
  doi:10.1103/PhysRevD.97.045013
  [arXiv:1710.06118 [hep-th]].
\bibitem{4} 
  E.~T.~Akhmedov, U.~Moschella and F.~K.~Popov,
  Phys.\ Rev.\ D {\bf 99}, no. 8, 086009 (2019)
  doi:10.1103/PhysRevD.99.086009
  [arXiv:1901.07293 [hep-th]].
\bibitem{5} 
  E.~T.~Akhmedov, U.~Moschella, K.~E.~Pavlenko and F.~K.~Popov,
  Phys.\ Rev.\ D {\bf 96}, no. 2, 025002 (2017)
  doi:10.1103/PhysRevD.96.025002
  [arXiv:1701.07226 [hep-th]].
\bibitem{6} 
  E.~T.~Akhmedov, N.~Astrakhantsev and F.~K.~Popov,
  JHEP {\bf 1409}, 071 (2014)
  doi:10.1007/JHEP09(2014)071
  [arXiv:1405.5285 [hep-th]].
\bibitem{7} 
  E.~T.~Akhmedov and F.~K.~Popov,
  JHEP {\bf 1509}, 085 (2015)
  doi:10.1007/JHEP09(2015)085
  [arXiv:1412.1554 [hep-th]].

\bibitem{8} 
  E.~T.~Akhmedov, H.~Godazgar and F.~K.~Popov,
  Phys.\ Rev.\ D {\bf 93}, no. 2, 024029 (2016)
  doi:10.1103/PhysRevD.93.024029
  [arXiv:1508.07500 [hep-th]].
\bibitem{9} 
  E.~T.~Akhmedov and S.~O.~Alexeev,
  Phys.\ Rev.\ D {\bf 96}, no. 6, 065001 (2017)
  doi:10.1103/PhysRevD.96.065001
  [arXiv:1707.02242 [hep-th]].

\bibitem{10} 
  L.~Astrakhantsev and O.~Diatlyk,
  International Journal of Modern Physics A 1 Vol. 33 (2018)
  doi:10.1142/S0217751X18501269
  [arXiv:1805.00549 [hep-th]].
\bibitem{11} 
  D.~A.~Trunin,
  Int.\ J.\ Mod.\ Phys.\ A {\bf 33}, no. 24, 1850140 (2018)
  doi:10.1142/S0217751X18501403
  [arXiv:1805.04856 [hep-th]].
(2018) [arXiv:1805.04856 [hep-th]].
\bibitem{12} 
  E.~T.~Akhmedov, O.~Diatlyk and A.~G.~Semenov,
  arXiv:1909.12805 [hep-th].

\bibitem{new1} 
A.~Riotto and M.~S.~Sloth,
JCAP \textbf{04}, 030 (2008)
doi:10.1088/1475-7516/2008/04/030
[arXiv:0801.1845 [hep-ph]].

\bibitem{new2} 
T.~Altherr,
Phys. Lett. B \textbf{341}, 325-331 (1995)
doi:10.1016/0370-2693(94)01327-9
[arXiv:hep-ph/9407249 [hep-ph]].

\bibitem{new3}
J.~Serreau and R.~Parentani,
Phys. Rev. D \textbf{87}, 085012 (2013)
doi:10.1103/PhysRevD.87.085012
[arXiv:1302.3262 [hep-th]].

\bibitem{new4}
B.~Mihaila, F.~Cooper and J.~F.~Dawson,
Phys. Rev. D \textbf{63}, 096003 (2001)
doi:10.1103/PhysRevD.63.096003
[arXiv:hep-ph/0006254 [hep-ph]].
\bibitem{new5} 
A.~Youssef and D.~Kreimer,
Phys. Rev. D \textbf{89}, 124021 (2014)
doi:10.1103/PhysRevD.89.124021
[arXiv:1301.3205 [gr-qc]].

\bibitem{new6} 
D.~Boyanovsky, H.~J.~de Vega, R.~Holman and M.~Simionato,
Phys. Rev. D \textbf{60}, 065003 (1999)
doi:10.1103/PhysRevD.60.065003
[arXiv:hep-ph/9809346 [hep-ph]].

\bibitem{new7} 
F.~Gautier and J.~Serreau,
Phys. Lett. B \textbf{727}, 541-547 (2013)
doi:10.1016/j.physletb.2013.10.072
[arXiv:1305.5705 [hep-th]].



\bibitem{46}
  E.~T.~Akhmedov, E.~N.~Lanina and D.~A.~Trunin,
  Phys.\ Rev.\ D {\bf 101}, no. 2, 025005 (2020)
  doi:10.1103/PhysRevD.101.025005
  [arXiv:1911.06518 [hep-th]].

\bibitem{14} 
  J.~S.~Schwinger,
  Phys.\ Rev.\  {\bf 82}, 664 (1951).
  doi:10.1103/PhysRev.82.664
\bibitem{15} 
  L.~V.~Keldysh,
  Zh.\ Eksp.\ Teor.\ Fiz.\  {\bf 47}, 1515 (1964)
  [Sov.\ Phys.\ JETP {\bf 20}, 1018 (1965)].

\bibitem{17} 
  J.~Berges,
  AIP Conf.\ Proc.\  {\bf 739}, no. 1, 3 (2004)
  doi:10.1063/1.1843591
  [hep-ph/0409233].
\bibitem{18} 
  J.~Rammer,
  Cambridge, UK: Univ. Pr. (2007) 536 p
\bibitem{19} 
  E.~A.~Calzetta and B.~L.~B.~Hu,
  doi:10.1017/CBO9780511535123


\bibitem{101}J. Spanier, K. B. Oldham, An Atlas of Functions, 361-372, Washington, DC: Hemisphere (1987)

\bibitem{Popov:1984mx}
V.~Popov,
``FUNCTIONAL INTEGRALS IN QUANTUM FIELD THEORY AND STATISTICAL PHYSICS,''
Published in: Dordrecht, Netherlands: Reidel ( 1983) 299 P. ( Mathematical Physics and Applied Mathematics, 8)
\bibitem{bb}
Erdelyi, A. (Editor) : Tables of Integral Transforms, Bateman Project
McGraw-Hill, New York 1954; Vol. 1.

\end{thebibliography}
\end{document}